\documentclass[aps,amsmath,amsfonts,nofootinbib,preprintnumbers,superscriptaddress,eqsecnum,secnumarabic]{revtex4}
\usepackage{graphicx,slashed,hyperref}
\pdfoutput=1
\usepackage[utf8]{inputenc}
\hypersetup{
   bookmarks=true,         
   unicode=true,          
   pdftoolbar=true,        
   pdfmenubar=true,        
   pdffitwindow=false,     
   pdfstartview={FitH},    
   pdftitle={My title},    
   pdfauthor={Author},     
   pdfsubject={Subject},   
   pdfcreator={Creator},   
   pdfproducer={Producer}, 
   pdfkeywords={keyword1} {key2} {key3}, 
   pdfnewwindow=true,      
   colorlinks=true,       
   linkcolor=blue,          
   citecolor=magenta,        
   filecolor=magenta,      
   urlcolor=blue           
}
\usepackage{amsmath}
\usepackage{amssymb}
\usepackage{comment}
\usepackage{latexsym,amsmath,amssymb,amsfonts}
\usepackage{graphicx}
\usepackage{array}
\usepackage{mathrsfs}
\usepackage{hyperref}
\usepackage{float}
\usepackage{wrapfig}
\usepackage{pdfpages}
\usepackage{fancyhdr}
\usepackage{wrapfig}
\usepackage{epsfig,verbatim,xspace,multirow,mathtools}
\usepackage{comment}
\usepackage{aas_macros}

\begin{document}

\title[Primordial black holes]{A brief review on primordial black holes as dark matter\footnote{Invited contribution to the journal \textit{Frontiers in Astronomy and Space Sciences - section Cosmology}, within the research topic \textit{When Planck, Einstein and Vera Rubin Meet. Dark Matter: What is it? Where is it?}}} 



\author{Pablo Villanueva-Domingo}
\email{pablo.villanueva.domingo@gmail.com}
\author{Olga Mena}
\email{olga.mena@ific.uv.es}
\author{Sergio Palomares-Ruiz}
\email{sergiopr@ific.uv.es}
\affiliation{Instituto de F\'isica Corpuscular (IFIC),
  CSIC-Universitat de Valencia,\\
Apartado de Correos 22085,  E-46071, Spain}

\begin{abstract}

Primordial black holes (PBHs) represent a natural candidate for one of the components of the dark matter (DM) in the Universe. In this review, we shall discuss the basics of their formation, abundance and signatures. Some of their characteristic signals are examined, such as the emission of particles due to Hawking evaporation and the accretion of the surrounding matter, effects which could leave an impact in the evolution of the Universe and the formation of structures. The most relevant probes capable of constraining their masses and population are discussed.

\medskip


\end{abstract}

\maketitle

\section{Introduction}

\label{sec:intro}

The hypothesis of the formation of black holes (BHs) in the early Universe was first suggested in 1967 by Zeldovich and Novikov \citep{1967SvA....10..602Z}, and independently by Hawking in 1971 \citep{1971MNRAS.152...75H}. Soon after, the possibility that primordial black holes (PBHs) could account for at least part of the DM became obvious \citep{1975Natur.253..251C, 1975A&A....38....5M}. At that time, the DM question started to be outlined as one of the fundamental problems in cosmology (see,  e.g., the reviews in \cite{2010dmp..book.....S, 2018RvMP...90d5002B}). DM (partly) composed by PBHs constitutes an exciting possibility, presenting an enormous number of observable signatures which can constrain its parameter space, as shall be detailed in Sec. \ref{sec:constraintsPBH}. The variety of phenomenological effects produced by PBHs allows placing stringent bounds on the abundance of PBHs, usually indicated by the energy fraction of DM as PBHs, $f_{\rm PBH}=\Omega_{\rm PBH}/\Omega_{\rm DM}$, with $\Omega_{\rm PBH}$ and $\Omega_{\rm DM}$ the ratios of energy densities and critical density. Moreover, since PBHs are usually expected to be formed before nucleosynthesis, BBN constraints on the baryon abundance do not apply to them, as they do not intervene in the nucleosynthesis of elements, and thus, can be regarded as non-baryonic DM \citep{2020ARNPS..7050520C}. Several recent reviews are devoted to discuss PBHs in great detail (see, e.g., \cite{Sasaki:2018dmp, Green:2020jor, 2020ARNPS..7050520C, Carr:2020gox}).

Shortly after the first detection of gravitational waves from a merger of $\sim 30 \, M_\odot$ BHs by LIGO \citep{Abbott:2016blz}, the question whether these could be of primordial nature was raised \citep{Bird:2016dcv}. Analysis of posterior data from the gravitational wave detectors LIGO and Virgo showed that the detected mergers are compatible with the hypothesis of their components being of primordial nature, although there is no strong preference over stellar BHs \citep{Clesse:2016vqa, Sasaki:2016jop, 2017PhRvD..96l3523A, 2018PDU....22..137C, DeLuca:2020qqa,   2020arXiv201013811G, Wong:2020yig, DeLuca:2021wjr}. PBHs with a lognormal mass function have been claimed to better fit data than BHs from astrophysical origin \citep{Dolgov:2020xzo}, although this is in contrast to the results of  \cite{Hall:2020daa}, and a mixed population seems compatible or even favored \citep{Hutsi:2020sol}. 

Unlike stellar BHs, formed from the collapse of a massive star, which can present masses only above $\sim 3 M_\odot$ (the Tolman–Oppenheimer–Volkoff limit \citep{1939PhRv...55..364T, 1939PhRv...55..374O}), PBHs could be produced with any mass. Thus, a positive measurement of a BH with a mass lower than $\sim 3 M_\odot$ would be a confirmation of the existence of primordial,  non-stellar BHs \citep{2018PDU....22..137C}. PBHs could also conform intermediate mass BHs, with masses between $\sim 10^2~M_\odot$ and $10^5~M_\odot$, too massive to be originated from a single star. It is the case of the merger event of BHs with masses $\sim 60~M_\odot$ and $\sim 80 M_\odot$, producing a remnant BH of $\sim 150 \, M_\odot$, in a so far mostly unobserved range of masses \citep{Abbott:2020tfl}. Furthermore, PBHs could constitute the seeds for super massive black holes (SMBHs), present in the nuclei of most galaxies, with masses ranging from $10^5~M_\odot$ to $10^{10} \, M_\odot$, and already existing at redshifts $z > 6$ \citep{2018MNRAS.478.3756C}. Such massive objects can hardly be produced from accreting stellar remnant BHs \citep{Volonteri_2010}. However, the existence of massive enough PBHs may act as seeds for the SMBHs, from which they could have grown by accretion.

PBH formation is already present in standard cosmologies, although extremely unlikely. However, their production usually requires some exotic inflationary scenarios or physics beyond the Standard Model (BSM) in order to obtain a large enough abundance. The typically considered formation mechanism of PBHs arises from the direct collapse of primordial fluctuations, whose power is enhanced at small scales as a consequence of some inflationary potential, as we shall further comment on below. There are, however, other scenarios which naturally predict a population of PBHs as a result of phase transitions in the early Universe and by the collapse of topological defects \citep{2016JCAP...02..064G, 2017JCAP...04..050D, Hawking:1987bn, PhysRevD.43.1106, PhysRevD.26.2681, 2017JCAP...12..044D}. Hence, the existence of PBHs would provide valuable hints about the still unknown physics of the very early Universe (see, e.g., \cite{Polnarev:1986bi, Khlopov:2008qy, Belotsky:2014kca}), and may allow to probe high-energy scales and supersymmetric theories \citep{Ketov:2019mfc}.

In this review, the most relevant aspects of PBHs as DM are briefly discussed, such as the mechanism of formation in section~\ref{sec:formation}, the initial abundance and mass distribution in section~\ref{sec:pbh_massfunc}, the process of accretion in section~\ref{sec:acc} and other PBH features, which could leave imprints on different observables, in section~\ref{sec:clusspin}. Current observational constraints on their population are summarized in section~\ref{sec:constraintsPBH}, and we conclude in section~\ref{sec:discus}.

\section{Formation and conditions of collapse}
\label{sec:formation}

\begin{figure}
\begin{center}
\includegraphics[scale=0.28]{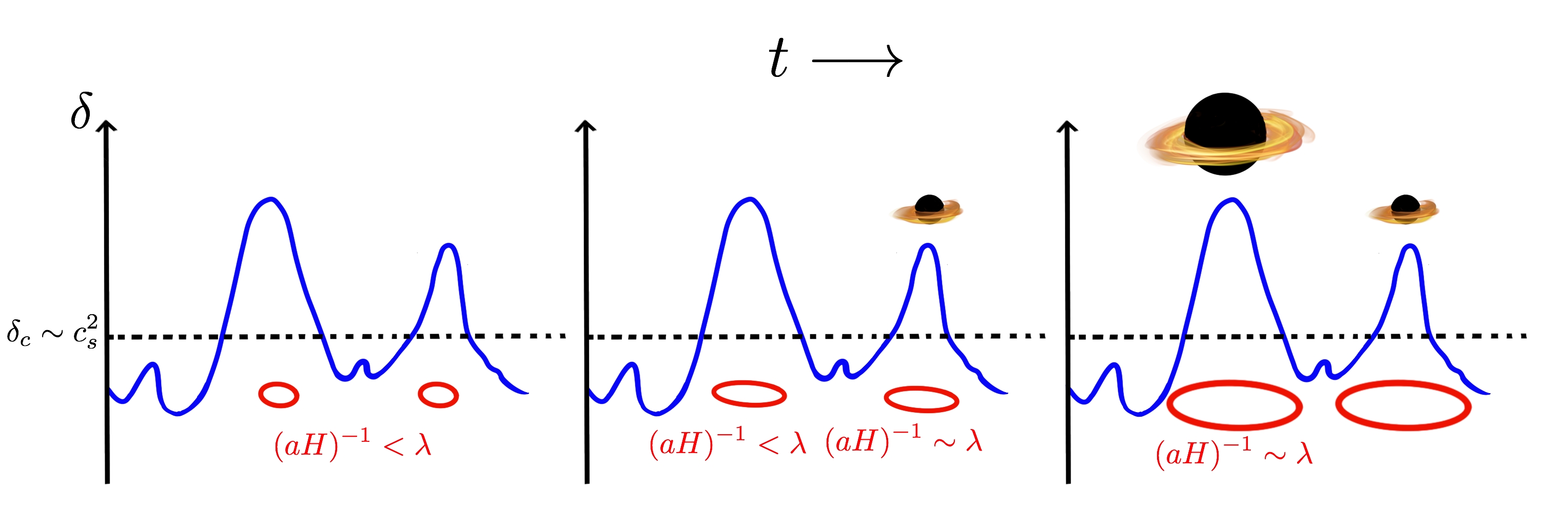}
\caption{Sketch of the formation of PBHs from overdensities for three different successive moments. When fluctuations larger than a critical threshold $\delta_c \sim c_s^2$ enter the horizon, i.e., their wavelength $\lambda=2\pi/k$ (which characterizes the size of the perturbation) is of the order of the Hubble horizon $(aH)^{-1}$, the overdense region collapses and a PBH is produced. As can be seen in Eqs.~(\ref{eq:MH}) and~(\ref{eq:Mk}), longer modes (large $\lambda$, low $k$) enter the horizon later and lead to more massive PBHs.}
\label{fig:pbhformation}
\end{center}
\end{figure}

The mass of a BH which collapsed in the early Universe depends on its formation time. A BH can be characterized by an extremely dense amount of matter in a very compact region, i.e., within the well-known Schwarzschild radius, $R_S = 2GM_{\rm PBH}/c^2 \sim 3 M_{\rm PBH}/M_\odot$ km. Thus, the mean density inside that region can be estimated as $\rho_S = M_{\rm PBH}/(4\pi R_S^3/3) \sim 10^{18} (M/M_\odot)^{-2}$ g cm$^{-3}$. On the other hand, the mean density of the universe in the radiation era scales with time as $\rho_c \sim 10^6 (t/{\rm s})^{-2}$ g cm$^{-3}$. In order to have PBH formation, densities at least of the order of the mean inside the BH horizon, $\rho_c \sim \rho_S$, are needed. Therefore, the mass of the resulting PBHs should be of the order of the horizon mass at that time, i.e., the mass within a region of the size of the Hubble horizon, $M_{\rm PBH} \sim M_{\rm H}$ \citep{1975ApJ...201....1C}, which is defined as
\begin{equation}
M_{\rm H} = \frac{4}{3} \, \bar{\rho}\left(\frac{c}{H}\right)^3 = \frac{c^3}{2 \, G H} \sim 10^{15} \; {\rm g}  \; \left( \frac{t}{10^{-23} {\rm s}} \right) ~.
\label{eq:MH}
\end{equation}
PBHs with masses of $\sim M_\odot \simeq 2 \times 10^{33}$~g would have been formed at around the QCD phase transition, at $t \sim 10^{-6}$ s. Since the PBH mass is roughly given by the mass within the horizon, it means that fluctuations entering the horizon can collapse into PBHs. A detailed calculation shows that $M_{\rm PBH} = \gamma M_{\rm H}$, where the proportionality factor $\gamma$ depends on the details of gravitational collapse, and gets values lower than 1. Early estimates showed that it can be approximated as $\gamma \simeq c_s^{3/2} \simeq 0.2$, with $c_s = 1/3$ the sound speed at the radiation epoch \citep{1975ApJ...201....1C}. More refined results show that the PBH mass is given by a scaling relation with the overdensity $\delta$, $M_{\rm PBH} = \kappa (\delta-\delta_c)^\alpha$ \citep{Niemeyer:1997mt, Niemeyer:1999ak}, where $\kappa$ and $\alpha$ are order unity constants, which depend on the background cosmology and on the shape of the perturbation \citep{Niemeyer:1997mt, Niemeyer:1999ak, Musco:2004ak}, and $\delta_c \simeq c_s^2$ is the collapse threshold (see, e.g. \cite{Escriva:2020tak, Musco:2020jjb} for recent accurate computations). Figure \ref{fig:pbhformation} depicts a sketch of the process of PBH formation.

Since PBHs are formed when fluctuations cross the horizon by the time of formation, $t_f$, their mass can be related to the wavelength of perturbations. When the mode of wavenumber $k$ crosses the horizon, the condition $a(t_f) \, H(t_{f}) = k$ holds. Since the mass of PBHs is proportional to the horizon mass at the moment of formation, $M_{\rm PBH} \propto \gamma H^{-1}$, at the radiation dominated era \citep{Sasaki:2018dmp},
\begin{equation}
M_{\rm PBH} \simeq 30\;  M_\odot \; \left( \frac{\gamma}{0.2} \right) \left( \frac{2.9 \times 10^5 {\rm Mpc}}{k} \right)^2 ~.
\label{eq:Mk}
\end{equation}
Hence, probing a given scale $k$ could constrain a PBH population of its corresponding mass.  Furthermore, an enhancement in the power spectrum around that scale would result in a large number of PBHs of such masses. In this review, we shall only consider PBHs formed during the radiation era. Those produced  before inflation ends would have been diluted due to their negligible density during the inflationary accelerated expansion. PBHs formed during the matter-dominated era, or in an early matter domination era previous to the radiation era, have also been considered in the literature, and may have different imprints, since the conditions of collapse are less restrictive, and could start from smaller inhomogeneities (see, e.g., \cite{Green:2020jor}).

\section{Abundance and mass function of PBHs}
\label{sec:pbh_massfunc}

It is possible to estimate the initial abundance of PBHs at the moment of formation, taking into account all overdensities above the threshold for collapse, $\delta_c \simeq 1/3$. Assuming a gaussian probability distribution, $P(\delta)$, for the overdensities with variance $\sigma^2(M)$ at a mass scale $M$, the initial abundance, defined as $\beta(M_{\rm H}) = \rho_{\rm PBH}(t_f)/\rho_{\rm tot}(t_f)$, is given by \citep{Sasaki:2018dmp, Green:2020jor}
\begin{equation}
\beta(M_{\rm PBH}) \simeq \gamma \int_{\delta_c}^\infty P(\delta) \, d\delta \simeq \gamma \, \sqrt{\frac{2}{\pi}} \, \frac{\sigma(M_{\rm PBH})}{\delta_c} \exp \left( -\frac{\delta_c^2}{2 \, \sigma^2(M_{\rm PBH})} \right) ~,
\end{equation}
where in the last equality, $\sigma(M_{\rm H}) \ll \delta_c$ has been assumed. For the standard cosmological scenario with an initial scale invariant power spectrum, at CMB scales, the amplitude of the fluctuations is around $\sigma(M_{\rm PBH}) \sim 10^{-5}$, leading to $\beta \sim 10^{-5} \exp(-10^{10})$, which is completely negligible \citep{2014arXiv1403.1198G}. Therefore, in order to have a relevant population of PBHs, larger values of the initial power spectrum are needed. On the other hand, the assumption of a gaussian distribution may not be consistent with enhanced fluctuations and the presence of PBHs, so deviations are unavoidable \citep{2018JCAP...03..016F, 2019JCAP...07..048D} (except for specific inflation models presenting an inflection point \cite{Atal:2018neu}). Non-gaussianities could have a great impact on the initial fraction and lead to a larger population, as well as leaving detectable signatures in gravitational waves \citep{Cai:2018dig}. Finally, note that the above formula follows the simple Press-Schechter approach, whereas to account for the non-universal nature of the threshold, the use of peak theory provides more accurate results \citep{Green:2004wb, Germani:2018jgr}. Its validity, however, is limited to relatively narrow power spectra, while broader spectra require the use of non-linear statistics \citep{Germani:2019zez}.

Nonetheless, although the initial fraction $\beta$ is a very small quantity, since matter and radiation densities scale differently with redshift ($\propto (1+z)^3$ and $\propto (1+z)^4$ respectively), the PBH contribution can become relevant at current times. The fraction $\beta(M)$ can be related to the current density parameters of PBH and radiation, $\Omega_{\rm PBH}$ and $\Omega_\gamma$, as \citep{2010PhRvD..81j4019C},
\begin{equation}
\Omega_{\rm PBH}(M) = \beta(M) (1+z_f) \Omega_\gamma \simeq \gamma^{1/2} \left( \frac{\beta(M)}{1.15 \times 10^{-8}} \right) \left( \frac{M}{M_\odot} \right)^{-1/2} ~.
\label{eq:abundanceomega}
\end{equation}
For initial fractions as low as $\beta \sim 10^{-8}$ of solar mass BHs, the fraction of energy in PBHs could be, thus, of order unity.

Depending on the specific mechanism of formation, a population of PBHs with different masses could be generated. The specific shape of the enhancement of fluctuations determines the mass distribution function. Sharp peaks in the power spectrum imply approximately \textit{monochromatic} distributions. For instance, chaotic new inflation may give rise to relatively narrow peaks \citep{PhysRevD.58.083510, 2008JCAP...06..024S}. However, inflation models with an inflection point in a plateau of the potential \citep{2017PDU....18...47G, 2017JPhCS.840a2032G}, or hybrid inflation \citep{2015PhRvD..92b3524C}, predict, instead, \textit{extended} mass functions, which can span over a large range of PBHs masses. In this review, we focus on monochromatic distributions for simplicity, although it is possible to translate constraints to extended mass functions, which can be very stringent despite the fact of having more parameters to fit \citep{2017PhRvD..96b3514C, 2018JCAP...01..004B}. Nonetheless, even if there are bounds that exclude $f_{\rm PBH} \sim 1$ in the monochromatic case, there are choices for the mass function which allow PBHs to constitute most or all of the DM abundance \citep{Lehmann:2018ejc, Sureda:2020vgi, Ashoorioon:2020hln}.

\section{Accretion onto PBHs}
\label{sec:acc}

One of the consequences of the existence of PBHs with greater impact on different observables is the process of accretion. Infalling matter onto PBHs would release radiation, injecting energy into the surrounding medium, and strongly impacting its thermal state, leaving significant observable signatures. The physics of accretion is highly complex, but one can attempt a simplified approach considering the spherical non-relativistic limit, following the seminal work by \cite{1952MNRAS.112..195B}. In this framework, the BH is treated as a point mass surrounded by matter, embedded in a medium which tends to constant density far enough from the BH. The relevant scale is the so-called Bondi radius, defined by
\begin{equation}
\label{eq:rBndi}
r_{\rm B} = \frac{G M_{\rm PBH}}{ c_{s,\infty}^2} ~, 
\end{equation}
where $c_{s,\infty}$ is the speed of sound far enough from the PBH. At distances around $ \sim r_{\rm B}$, the accretion process starts to become relevant, and the velocity of the accreted matter reaches $v \sim c_{s, \infty}$, when the density is still close to the boundary value, $\rho_\infty$. Taking into account the velocity of the BH relative to the medium, $v_{\rm rel}$, one can write the accretion rate as \citep{1952MNRAS.112..195B}
\begin{equation}
\dot{M}_{\rm PBH} = 4\pi \, \lambda \, \rho_\infty \, \frac{(G M_{\rm PBH})^2}{(c_{s,\infty}^2 + v_{\rm rel}^2)^{3/2}} ~,
\label{eq:mdotvlin}
\end{equation}
where the parameter $\lambda$ is the dimensionless accretion rate and depends on the equation of state, but it is order $\sim 1$ for the cases of interest \citep{Ali-Haimoud:2016mbv}. In the early Universe, $v_{\rm rel}$ can be estimated as the baryon-DM relative velocity, computed in linear theory. Its root-mean-square value is approximately constant before recombination, dropping linearly with $1+z$ at later times \citep{Ali-Haimoud:2016mbv}.

On the other hand, real BHs spin and form an accreting disk. Thus, the spherical symmetric case may not be applicable. Even though PBH spins are expected to be small, an accretion disk would form if the angular momentum is large enough to keep matter orbiting at Keplerian orbits at distances much larger than the innermost stable orbits, which are roughy given by the Schwarzschild radius \citep{Agol:2001hb}. Applying this criterion, the formation of accretion disks around PBHs has been suggested to occur if the condition $f_{\rm PBH} M_{\rm PBH}/M_\odot \ll ((1+z)/730)^3$ is fulfilled \citep{Poulin:2017bwe}. This is satisified for $M_{\rm PBH} \gtrsim M_\odot$ and large enough abundances at the epoch of CMB decoupling or at later times. Some of the results outlined above are still valid if the dimensionless accretion rate $\lambda$ is suppresed by roughly two orders of magnitude, after accounting for viscosity effects and matter outflows through jets \citep{Poulin:2017bwe}.

The matter falling onto BHs is greatly accelerated, which gives rise to radiative emission of high-energy photons by processes such as bremsstrahlung. The luminosity of accreting BHs is proportional to its accretion rate, and can be written as \citep{Xie:2012rs}
\begin{equation}
L_{\rm acc} = \epsilon(\dot{M}_{\rm PBH}) \, \dot{M}_{\rm PBH} ~,
\label{eq:lacc}
\end{equation}
where $\epsilon(\dot{M}_{\rm PBH}) $ denotes the radiative efficiency, which is in general a complicated function of the accretion rate $\dot{M}_{\rm PBH}$, and depends upon the geometry, viscosity and other hydrodynamical considerations.

If the accretion disk is optically thin, most of the energy released through viscous dissipation is radiated away, and the luminosities obtained can be close to the Eddington luminosity, $L_{\rm Edd}$, explaining the extreme brightness of many far Active Galactic Nuclei (AGN) \citep{Shakura:1972te}. However, PBHs, like nearby astrophysical BHs, may radiate in a much less efficient way, through the Advective-Dominated-Accretion-Flow (ADAF) \citep{Ichimaru:1977uf, Rees:1982pe, 1994ApJ...428L..13N, Narayan:1994is} (see, e. g., \cite{Yuan:2014gma} for a review). In this scenario, the dynamics is ruled by advective currents, forming a hot thick disk. Most of the emitted energy is deposited in the same accretion disk, heating it up. Thus, only a small portion of energy is released to the surrounding medium, the radiative process being inefficient. In the ADAF scenario, the efficiency function can be fitted by a broken power-law formula, with the slopes and amplitudes dependent on the mass range and the specific modeling of viscosity effects \citep{Xie:2012rs}.

Finally, the energy emitted in the accretion processes is deposited through different channels into the medium. The energy deposition rate for each channel reads \citep{Ali-Haimoud:2016mbv, Poulin:2017bwe}
\begin{equation}
\label{eq:eninj}
\left(\frac{dE_c}{dV dt}\right)_{{\rm dep}} = f_c(z) \,L_{\rm acc} \, n_{\rm PBH} = f_c(z) \, L_{\rm acc} \, \frac{f_{\rm PB
H} \, \rho_{\rm DM}}{M_{\rm PBH}} ~,
\end{equation}
where the subscript $c$ denotes the channel in which energy is deposited, namely: ionization, heating, or atomic excitations (where the ${\rm Ly}\alpha$ transitions are the most relevant ones). The energy deposition factors $f_c(z)$ quantify the fraction of energy which goes to the different channels, and has been computed numerically \citep{Slatyer:2015kla}. This energy injection would affect different types of observables, as we briefly outline below.

\section{Other generic features}
\label{sec:clusspin}

A noteworthy phenomenon of BHs is that of evaporation. Due to quantum effects in curved spacetimes, BHs may emit particles at their event horizon, as was noticed in \cite{1974Natur.248...30H}. The emitted radiation would have a nearly thermal black body spectrum, with a temperature given by \citep{2010PhRvD..81j4019C}
\begin{equation}
	T_{\rm BH} = \frac{\hbar c^3}{8\pi k_B GM} \sim 10^{-7} \; {\rm K} \; \frac{M_\odot}{M} ~,
	\label{eq:HawkingT}
\end{equation}
which is known as Hawking temperature. Due to this emission, BHs would slowly lose mass until completely evaporate. The lifetime of a PBH of initial mass $M$ is \citep{2003PhTea..41..299L}
\begin{equation}
	\tau(M) \sim 10^{64} \; {\rm yr} \; \left( \frac{M}{M_\odot} \right)^3 ~.
\end{equation}
Thus, the lower the PBH mass, the earlier it evaporates. Those with masses of $10^{15}$~g or below would have already evaporated by now, having lifetimes shorter than the age of the Universe \citep{1976PhRvD..14.3260P}, so they cannot contribute to the current DM abundance. These evaporation products or the effects they produce in different observables can be search for in a variety of experiments, probing different mass ranges. Detailed computations of the emitted spectra can be performed by codes such as {\tt BlackHawk} \cite{Arbey:2019mbc}.

Another important feature is that of clustering. If fluctuations are originally Gaussian distributed and around a relatively narrow peak, PBHs are not expected to be originated in clusters, being initially randomly distributed on small scales \citep{2018PhRvL.121h1304A, 2018PhRvD..98l3533D}. However, either primordial non-gaussianities or a broad peak in the power spectrum could lead to a significant initial clustering \citep{2019PTEP.2019j3E02S, 2018JCAP...10..043B} (although broad spectra have also been argued not to produce appreciable clustering \citep{2019JCAP...11..001M}). Anyway, PBHs could become bounded as the Universe evolves. A proper determination of their clustering properties at later times is of great importance, for instance, in order to estimate their merger rates \citep{2018PhRvD..98l3533D, DeLuca:2020jug}. Indeed, the formation of clusters could alleviate some constraints on the PBH abundance \citep{Belotsky:2018wph}.

On another hand, since PBHs would be formed from the collapse of high density peaks relatively spherically symmetric, their torques and angular momentum are expected to be small \citep{Chiba:2017rvs, Mirbabayi:2019uph}. It is usually quantified with the dimensionless spin parameter, $\mathcal{S} = S/(GM_{\rm PBH}^2)$, where $S$ is the spin. Estimations of $\mathcal{S}$ for PBHs show that it is a small quantity, equal or lower than $0.01$ \citep{2019JCAP...05..018D}. In contrast, astrophysical BHs are expected to have substantially larger spins, since angular momentum must be conserved during the collapse of their stars of origin, which are often rotating. Hence, the spin can serve as a good proxy to distinguish the nature of a population of BHs. The measurement of low spin parameters could represent a hint for the detection of PBHs. The latest Bayesian analyses of LIGO/Virgo mergers suggest that low values of the spin parameter are strongly preferred by data, regardless of the assumed priors \citep{2020arXiv201013811G}. Note, however, that the PBH mass and spin depend on the accretion mechanism and their time evolution is correlated \citep{DeLuca:2020bjf}.

Furthermore, due to the discrete nature of PBHs, a Poisson shot noise contribution to the matter power spectrum, constant in wavenumber, $P_{\rm sn}(k) \propto f_{\rm PBH}^2 \, \bar{n}_{\rm PBH}^{-1} \propto f_{\rm PBH} \, M_{\rm PBH}$, would be expected \citep{2003ApJ...594L..71A, Gong:2017sie}. PBHs fluctuations give rise to isocurvature modes \citep{2003ApJ...594L..71A, 2006PhRvD..73h3504C, Inman:2019wvr}, and thus, affect only at scales smaller than the horizon at the epoch of matter-radiation equality \citep{Peacock:1999ye}. Therefore, this leads to an enhancement in the matter power spectrum, increasing the population of low-mass halos, which can be constrained by large scale structure and Ly$\alpha$ forest analyses. This effect  is different from the one induced by other non-CDM candidates, such as warm DM or fuzzy DM, which suppress small scale fluctuations, washing out small structures. This contribution becomes relevant for low-mass halos not large enough to cool and collapse to form stars, which are commonly known as \textit{minihalos}. It has been argued that this enhancement may produce a non-negligible 21 cm signal from the neutral hydrogen in minihalos \citep{Gong:2017sie, Gong:2018sos}. However, consistently accounting for the heating of the IGM due to PBH accretion increases the Jeans mass and suppresses the minihalo 21 cm signal, making it almost negligible \citep{Mena:2019nhm}.

\section{Observational constraints on PBHs as DM}
\label{sec:constraintsPBH}

\begin{figure}
\begin{center}
\includegraphics[scale=0.85]{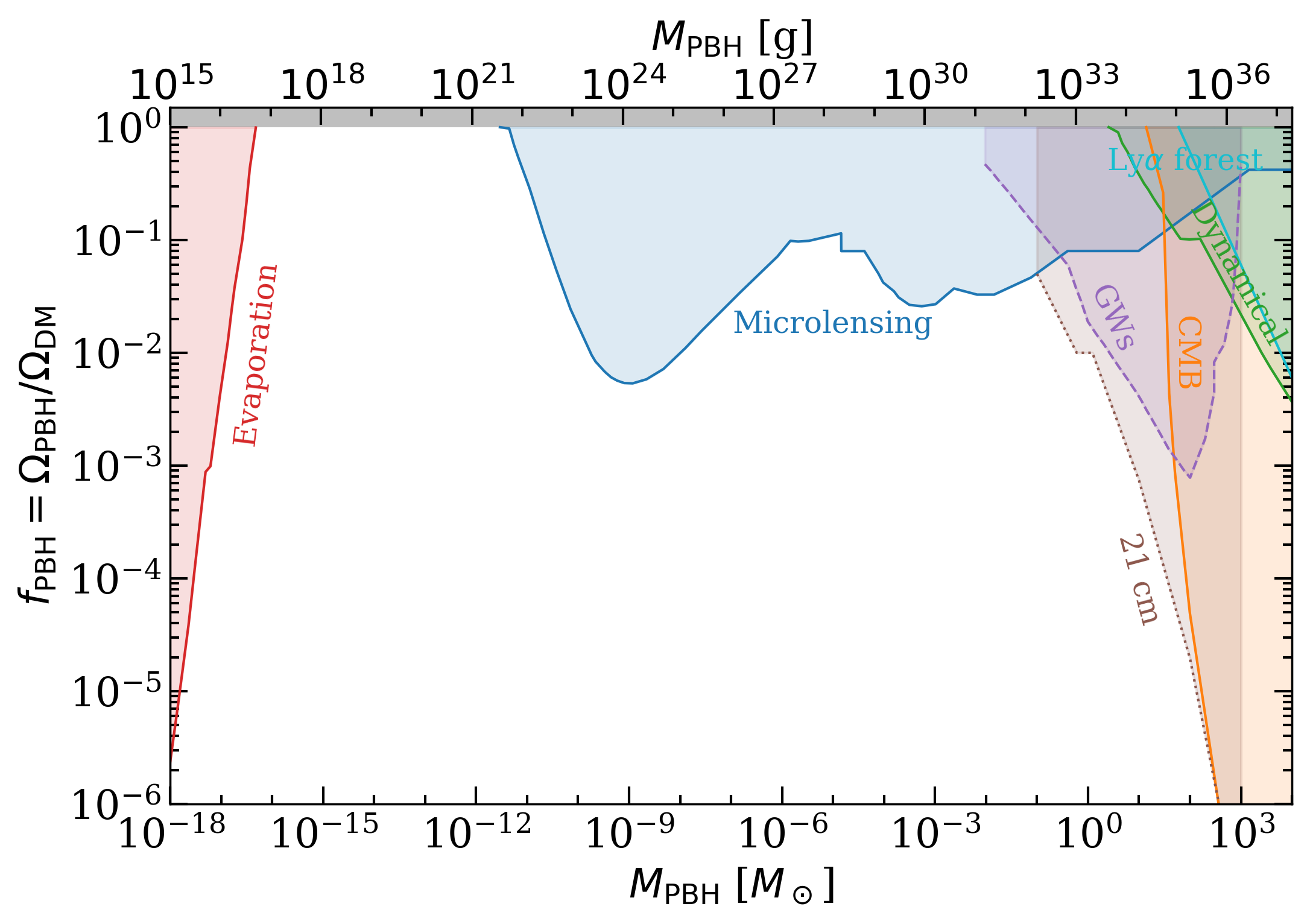}
\caption{Compilation of contraints on the PBH fraction (with respect to DM) as a function of the PBH mass, assuming a monochromatic mass function. The different probes considered are: impact of PBH evaporation (red) on the extragalactic $\gamma$-ray background \citep{2010PhRvD..81j4019C} and on the CMB spectrum \citep{Clark:2016nst}; non-observation of microlensing events (blue) from the MACHO \citep{Alcock:2000kg}, EROS \citep{Tisserand:2006zx}, Kepler \citep{Griest:2013aaa}, Icarus \citep{Oguri:2017ock}, OGLE \citep{2019PhRvD..99h3503N} and Subaru-HSC \citep{Croon:2020ouk} collaborations; PBH accretion signatures on the CMB (orange), assuming spherical accretion of PBHs within halos \citep{2020PhRvR...2b3204S}; dynamical constraints, such as disruption of stellar systems by the presence of PBHs (green), on wide binaries \citep{2014ApJ...790..159M} and on ultra-faint dwarf galaxies \citep{2016ApJ...824L..31B}; power spectrum from the Ly$\alpha$ forest (cyan) \citep{2019PhRvL.123g1102M}; merger rates from gravitational waves (purple), either from individual mergers \citep{2018PhRvD..98b3536K, Authors:2019qbw} or from searches of stochastic gravitational wave background \citep{2020JCAP...08..039C}. Gravitational waves limits are denoted by dashed lines, since they could be invalidated \citep{Boehm:2020jwd}. Dotted brown line corresponds to forecasts from the 21 cm power spectrum with SKA sensitivities \citep{Mena:2019nhm} and from 21 cm forest prospects \citep{Villanueva-Domingo:2021cgh}. Figure created with the publicly available \texttt{Python} code \href{https://github.com/bradkav/PBHbounds}{\texttt{PBHbounds}}  \citep{bradley_j_kavanagh_2019_3538999}.}
\label{fig:pbhbounds}
\end{center}
\end{figure}

PBHs can impact cosmology and astrophysics in a wide range of ways, leaving different observational effects which allow to constrain their properties. In this section, we review the most important bounds on the current fraction of PBHs as DM, $f_{\rm PBH} = \Omega_{\rm PBH}/\Omega_{\rm DM}$, for a wide range of masses $M_{\rm PBH}$, for monochromatic mass functions. A collection of limits from the different probes is depicted in Fig. \ref{fig:pbhbounds}. For a more comprehensive list of constraints, see \cite{Green:2020jor, 2020ARNPS..7050520C}.

\begin{itemize}

\item \textbf{Evaporation}

Since BHs emit energy due to Hawking radiation, those with a lifetime shorter than the age of the Universe must have disintegrated nowadays, a fact which excludes PBHs with $M_{\rm PBH}<M_* \simeq 4\times 10^{14}$~g to form part of the current DM \citep{1976PhRvD..14.3260P}. Moreover, PBHs with masses small enough, although still present, should emit a strong $\gamma$ ray and cosmic ray background which could be observed. Absence of its detection strongly constrains the range of masses $M_{\rm PBH} \lesssim 10^{17}$~g. In particular, the maximum fraction allowed is $f_{\rm PBH} \lesssim 2 \times 10^{-8} (M_{\rm PBH}/M_*)^{3+\epsilon}$, with $\epsilon \sim 0.1 - 0.4$ \citep{2016PhRvD..94h3504C}. Comparable limits have been found from INTEGRAL and COMPTEL observations of the Galactic Center \citep{DeRocco:2019fjq, Laha:2019ssq, Laha:2020ivk, Coogan:2020tuf}. Furthermore, bounds from the isotropic X-ray and soft gamma-ray background have also been recently updated \citep{Ballesteros:2019exr, Iguaz:2021irx}. Additionally, data on the diffuse supernova neutrino background at Super-Kamiokande are also able to set constraints \citep{Dasgupta:2019cae}.

\item \textbf{Microlensing}

If a compact object crosses the line of sight of a star, it may produce a so-called microlensing effect, which implies a transient and achromatic amplification of its flux. The range of masses of the objects which can produce it span from $5 \times 10^{-10}$ to $\sim 100 \, M_\odot$ \citep{1986ApJ...304....1P, Green:2020jor}. The non-detection of these events leads to bounds on the maximum abundance of PBHs about $f_{\rm PBH} \lesssim 0.01-0.1$ by the MACHO \citep{Alcock:2000kg} and EROS \citep{Tisserand:2006zx} surveys in the Large and Small Magellanic Clouds, the Subaru Hyper Suprime Cam (HSC) in M31 (Andromeda) \citep{2019NatAs...3..524N} and the Optical Gravitational Lensing Experiment (OGLE) in the Galactic bulge \citep{2019PhRvD..99h3503N}. Nonetheless, the existence of Earth-mass PBHs ($M_{\rm PBH} \sim 10^{-5} M_\odot$) with a fraction $f_{\rm PBH} \sim 0.03$ could explain the observed excess of 6 microlensing events found in the OGLE data \citep{2017Natur.548..183M}, which is consistent with other constraints in this range of PBHs masses \citep{2019PhRvD..99h3503N}. Although this may constitute a hint of their existence, it cannot be regarded as a detection of PBHs, since these microlensing observations could also be explained by free-floating planets. There are some caveats regarding the results of the MACHO collaboration \citep{2015A&A...575A.107H}, since the limits reported are model dependent and could be biased by the assumption of an over-massive halo. Moreover, the results of the MACHO and EROS projects have been found to be statistically incompatible. Therefore, these bounds are not completely reliable, and PBHs could not be definitely ruled out within this range of masses. Variations of the lensing effect in type Ia Supernovae \citep{2018PhRvL.121n1101Z} and gamma-ray bursts \citep{2012PhRvD..86d3001B} have also been proposed to constrain the PBH parameter space, although these limits have been later invalidated \citep{2017arXiv171206574G, 2018JCAP...12..005K}.

\item \textbf{Gravitational waves}

The observation of BH mergers by LIGO and Virgo collaborations can be employed to constrain the allowed number of PBHs. Demanding that the predicted merger rates of PBH binaries cannot exceed the ones measured by gravitational waves, tight upper bounds of $f_{\rm PBH} \lesssim 0.01$ have been found for PBHs masses between 1 and 300 $M_\odot$ \citep{2017PhRvD..96l3523A, 2018PhRvD..98b3536K}. The non-observation of a stochastic gravitational wave background of mergers expected from a population of PBHs has also been used to constrain their abundace \citep{2020JCAP...08..039C}. However, to derive these limits, BHs have been treated as Schwarzschild BHs, while it would be more appropiate to use cosmological BH solutions embedded in a FLRW metric, such as the Thakurta metric \citep{1981InJPh..55..304T}. This implies a time-dependent mass, and that PBH binaries created before galaxy formation would have merged at much earlier times, allowing to obtain merger rates consistent with the LIGO observations, and completely avoiding these constraints \citep{Boehm:2020jwd}.

\item \textbf{Dynamical constraints}

Due to two-body interactions, kinetic energies of systems of different masses usually become balanced and match. If a stellar system also has an additional MACHO population, its stars would gain kinetic energy and, due to the virial theorem, the system would expand. Therefore, the presence of PBHs would dynamically heat star clusters, making them larger and with higher velocity dispersions, leading to an eventual dissolution into its host galaxy. Populations with high mass to luminosity ratios are more sensitive to this effect, as happens with ultra faint dwarf galaxies (UFDW), which would be disrupted by the presence of PBHs. Making use of these effects, tight bounds have been obtained at $f_{\rm PBH} \sim 10^{-3}$ for $M_{\rm PBH} \sim 10^4 \, M_\odot$, weakening at lower masses down to $f_{\rm PBH} \lesssim 1$ for $M_{\rm PBH} \sim 10 \, M_\odot$ \citep{2016ApJ...824L..31B}. In a similar way, wide binary stellar systems may be perturbed by compact objects, potentially being disrupted after multiple encounters. The separation distribution of wide binaries restricts the PBH fraction from $f_{\rm PBH} \lesssim 1$ for $M_{\rm PBH} \simeq 3 \, M_\odot$ down to $f_{\rm PBH} \lesssim 0.1$ at $M_{\rm PBH} \gtrsim 70 \, M_\odot$ \citep{2014ApJ...790..159M}.

\item \textbf{CMB}

Radiation emitted either by accretion or from Hawking evaporation may affect the CMB spectrum in two ways: producing spectral distortions and modifying temperature anisotropies. The energetic radiation can enhance the ionization rate, delay recombination and shift the peaks of the CMB anisotropy spectrum, as well as induce more diffusion damping. The polarization spectrum could also be modified, since the increase of the free electron fraction would increase the Thomson optical depth and enhance the reionization bump at large angular scales. Although early CMB analyses \citep{2008ApJ...680..829R} found very stringent bounds on the allowed abundance of accreting PBHs, later works revisited these computations and found much milder constraints \citep{2016arXiv161207264H, Ali-Haimoud:2016mbv}. On another hand, while the former constraints rely on the assumption of spherical accretion, accreting disks have been argued to be more realistic for PBHs, resulting in tighter limits \citep{Poulin:2017bwe}. Taking into account that PBHs could be immersed in DM halos with higher densities than the background, their accretion rates would be increased, also leading to more stringent constraints \citep{2020PhRvR...2b3204S}. CMB limits from accretion are currently the most stringent ones for masses $\gtrsim 10 \, M_\odot$. The main caveat is their dependence on some details of the accretion mechanisms, such as the effective velocity and the accretion rate, which may not be very well understood yet. 

On the other hand, the energy injection from PBHs evaporation would produce anisotropies and spectral distortions in the CMB spectrum, which would also limit the maximum abundance, leading to similar costraints to those obtained from the extra-galactic $\gamma$-ray background commented above \citep{Poulin:2016anj, Clark:2016nst, Acharya:2019xla}. Besides energy injection from accretion of BH evaporation, spectral distortions can also be produced by other means, such as the diffusion of photons due to Silk damping at small scales. This allows the translation of constraints on spectral distortions from FIRAS to stringent upper bounds on the PBH abundance, for masses $M_{\rm PBH}>10^5M_\odot$ \citep{Nakama:2017xvq}.

\item \textbf{Ly$\alpha$ forest}

The discrete nature of PBHs would lead to a shot-noise contribution to the matter power spectrum, enchancing small scale fluctuations. Observations of the Ly$\alpha$ forest, which traces matter distribution at small scales, have been employed to extract limits on the maximum allowed fraction of PBHs \citep{2003ApJ...594L..71A}. The shot-noise contribution to the power spectrum depends on the joint product $f_{\rm PBH}M_{\rm PBH}$, for which the upper bound $f_{\rm PBH}M_{\rm PBH} \leq 60 \, M_\odot$ has been obtained \citep{2019PhRvL.123g1102M}. The drawback of this method is on the priors of the reionization modeling and, as any Ly$\alpha$ forest analysis, is model dependent.

\item \textbf{21 cm cosmology}

The 21 cm line signal from the hyperfine structure of the hydrogen is highly sensitive to the thermal state of the IGM, and thus, energy injection from PBH accretion or evaporation may leave strong observable signatures. The first claimed measurement of a global absorption dip by the EDGES collaboration \citep{Bowman:2018yin} may lead to competitive bounds on the PBH abundance, either from accretion processes \citep{Hektor:2018qqw} or from energy injection by evaporation \citep{Clark:2018ghm, Halder:2021jiv, Halder:2021rbq}. It must be noted, however, that the EDGES signal has not been confirmed yet by other experiments, and it has been argued that it could be explained by alternative mechanisms \citep{2018Natur.564E..32H, 2019ApJ...874..153B, 2020MNRAS.492...22S}. On the other hand, although the cosmological 21 cm power spectrum has not been detected yet, forecasts with future experiments such as HERA and SKA have shown that 21 cm power spectrum data from these two radiotelescopes could potentially improve the bounds up to $f_{\rm PBH}< 10^{-2}-10^{-6}$ for masses above $M_\odot$ \citep{Mena:2019nhm}. The 21 cm forest observed as absorption troughs in the spectra of radio sources at $z\sim 10-15$ could also provide similar limits on the PBH abundance, due to the Poisson shot noise and to the accretion heating effect \citep{Villanueva-Domingo:2021cgh}.

\end{itemize}

\section{Conclusions}
\label{sec:discus}

The extremely rich physics involved in the formation, evolution and distribution of PBHs implies a large number of observable effects which allow probing them. A myriad of constraints are present for a large range of PBHs masses. In recent years some of these limits, as those from microlensing, femtolensing, CMB accretion or BH mergers, have been revisited. Some of these bounds have been significantly weakened or have even disappeared, opening windows in the parameter space where PBHs could still form a substantial part of the DM, if not all. On the other hand, future experiments with better sensitivities may be able to reach yet unexplored regions in the parameter space and tighten up current limits. New probes, such as the 21 cm line pursued in radio interferometers like SKA, will present a promising and  powerful way to proof or refute the existence of PBHs formed in the early universe and their potential contribution to the DM in the Universe.

\section*{Acknowledgments}
PVD and OM are supported by the Spanish grants FPA2017-85985-P and PROMETEO/2019/083. SPR is supported by the Spanish FEDER/MCIU-AEI grant FPA2017-84543-P, and partially, by the Portuguese FCT (UID/FIS/00777/2019 and CERN/FIS-PAR/0004/2019). All authors acknowledge support from the European ITN project HIDDeN (H2020-MSCA-ITN-2019//860881-HIDDeN).

\bibliographystyle{JHEP}
\bibliography{biblioreview}

\providecommand{\href}[2]{#2}\begingroup\raggedright\begin{thebibliography}{100}

\bibitem{1967SvA....10..602Z}
Y.~B. {Zel'dovich} and I.~D. {Novikov}, \emph{{The hypothesis of cores retarded
  during expansion and the hot cosmological model}}, {\emph{\sovast} {\bf 10}
  (Feb., 1967) 602}.

\bibitem{1971MNRAS.152...75H}
S.~{Hawking}, \emph{{Gravitationally collapsed objects of very low mass}},
  \href{http://dx.doi.org/10.1093/mnras/152.1.75}{\emph{\mnras} {\bf 152}
  (Jan., 1971) 75}.

\bibitem{1975Natur.253..251C}
G.~F. {Chapline}, \emph{{Cosmological effects of primordial black holes}},
  \href{http://dx.doi.org/10.1038/253251a0}{\emph{\nat} {\bf 253} (Jan., 1975)
  251--252}.

\bibitem{1975A&A....38....5M}
P.~{Meszaros}, \emph{{Primeval black holes and galaxy formation.}},
  {\emph{\aap} {\bf 38} (Jan., 1975) 5--13}.

\bibitem{2010dmp..book.....S}
R.~H. {Sanders}, \emph{{The dark matter problem: a historical perspective}}.
\newblock 2010.

\bibitem{2018RvMP...90d5002B}
G.~{Bertone} and D.~{Hooper}, \emph{{History of dark matter}},
  \href{http://dx.doi.org/10.1103/RevModPhys.90.045002}{\emph{Reviews of Modern
  Physics} {\bf 90} (Oct., 2018) 045002},
  [\href{http://arxiv.org/abs/1605.04909}{{\tt 1605.04909}}].

\bibitem{2020ARNPS..7050520C}
B.~{Carr} and F.~{K{\"u}hnel}, \emph{{Primordial black holes as dark matter:
  recent developments}},
  \href{http://dx.doi.org/10.1146/annurev-nucl-050520-125911}{\emph{Annual
  Review of Nuclear and Particle Science} {\bf 70} (Oct., 2020) annurev},
  [\href{http://arxiv.org/abs/2006.02838}{{\tt 2006.02838}}].

\bibitem{Sasaki:2018dmp}
M.~Sasaki, T.~Suyama, T.~Tanaka and S.~Yokoyama, \emph{{Primordial black
  holes—perspectives in gravitational wave astronomy}},
  \href{http://dx.doi.org/10.1088/1361-6382/aaa7b4}{\emph{Class. Quant. Grav.}
  {\bf 35} (2018) 063001}, [\href{http://arxiv.org/abs/1801.05235}{{\tt
  1801.05235}}].

\bibitem{Green:2020jor}
A.~M. Green and B.~J. Kavanagh, \emph{{Primordial black holes as a dark matter
  candidate}}, \href{http://dx.doi.org/10.1088/1361-6471/abc534}{\emph{J. Phys.
  G} {\bf 48} (2021) 4}, [\href{http://arxiv.org/abs/2007.10722}{{\tt
  2007.10722}}].

\bibitem{Carr:2020gox}
B.~Carr, K.~Kohri, Y.~Sendouda and J.~Yokoyama, \emph{{Constraints on
  primordial black holes}},  \href{http://arxiv.org/abs/2002.12778}{{\tt
  2002.12778}}.

\bibitem{Abbott:2016blz}
{\scshape LIGO Scientific, Virgo} collaboration, B.~P. Abbott et~al.,
  \emph{{Observation of gravitational waves from a binary black hole merger}},
  \href{http://dx.doi.org/10.1103/PhysRevLett.116.061102}{\emph{Phys. Rev.
  Lett.} {\bf 116} (2016) 061102}, [\href{http://arxiv.org/abs/1602.03837}{{\tt
  1602.03837}}].

\bibitem{Bird:2016dcv}
S.~Bird et~al., \emph{{Did LIGO detect dark matter?}},
  \href{http://dx.doi.org/10.1103/PhysRevLett.116.201301}{\emph{Phys. Rev.
  Lett.} {\bf 116} (2016) 201301}, [\href{http://arxiv.org/abs/1603.00464}{{\tt
  1603.00464}}].

\bibitem{Clesse:2016vqa}
S.~Clesse and J.~Garc\'\i{}a-Bellido, \emph{{The clustering of massive
  primordial black holes as dark matter: measuring their mass distribution with
  Advanced LIGO}},
  \href{http://dx.doi.org/10.1016/j.dark.2016.10.002}{\emph{Phys. Dark Univ.}
  {\bf 15} (2017) 142--147}, [\href{http://arxiv.org/abs/1603.05234}{{\tt
  1603.05234}}].

\bibitem{Sasaki:2016jop}
M.~Sasaki, T.~Suyama, T.~Tanaka and S.~Yokoyama, \emph{{Primordial black hole
  scenario for the gravitational-wave wvent GW150914}},
  \href{http://dx.doi.org/10.1103/PhysRevLett.117.061101}{\emph{Phys. Rev.
  Lett.} {\bf 117} (2016) 061101}, [\href{http://arxiv.org/abs/1603.08338}{{\tt
  1603.08338}}].

\bibitem{2017PhRvD..96l3523A}
Y.~{Ali-Ha{\"\i}moud}, E.~D. {Kovetz} and M.~{Kamionkowski}, \emph{{Merger rate
  of primordial black-hole binaries}},
  \href{http://dx.doi.org/10.1103/PhysRevD.96.123523}{\emph{\prd} {\bf 96}
  (Dec., 2017) 123523}, [\href{http://arxiv.org/abs/1709.06576}{{\tt
  1709.06576}}].

\bibitem{2018PDU....22..137C}
S.~{Clesse} and J.~{Garc{\'\i}a-Bellido}, \emph{{Seven hints for primordial
  black hole dark matter}},
  \href{http://dx.doi.org/10.1016/j.dark.2018.08.004}{\emph{Physics of the Dark
  Universe} {\bf 22} (Dec., 2018) 137--146},
  [\href{http://arxiv.org/abs/1711.10458}{{\tt 1711.10458}}].

\bibitem{DeLuca:2020qqa}
V.~De~Luca, G.~Franciolini, P.~Pani and A.~Riotto, \emph{{Primordial black
  holes confront LIGO/Virgo data: current situation}},
  \href{http://dx.doi.org/10.1088/1475-7516/2020/06/044}{\emph{JCAP} {\bf 06}
  (2020) 044}, [\href{http://arxiv.org/abs/2005.05641}{{\tt 2005.05641}}].

\bibitem{2020arXiv201013811G}
J.~{Garcia-Bellido}, J.~F. {Nu{\~n}o Siles} and E.~{Ruiz Morales},
  \emph{{Bayesian analysis of the spin distribution of LIGO/Virgo black
  holes}}, {\emph{arXiv e-prints} (Oct., 2020) arXiv:2010.13811},
  [\href{http://arxiv.org/abs/2010.13811}{{\tt 2010.13811}}].

\bibitem{Wong:2020yig}
K.~W.~K. Wong et~al., \emph{{Constraining the primordial black hole scenario
  with Bayesian inference and machine learning: the GWTC-2 gravitational wave
  catalog}}, \href{http://dx.doi.org/10.1103/PhysRevD.103.023026}{\emph{Phys.
  Rev. D} {\bf 103} (2021) 023026},
  [\href{http://arxiv.org/abs/2011.01865}{{\tt 2011.01865}}].

\bibitem{DeLuca:2021wjr}
V.~De~Luca, G.~Franciolini, P.~Pani and A.~Riotto, \emph{{Bayesian evidence for
  both astrophysical and primordial black holes: mapping the GWTC-2 catalog to
  third-generation detectors}},  \href{http://arxiv.org/abs/2102.03809}{{\tt
  2102.03809}}.

\bibitem{Dolgov:2020xzo}
A.~D. Dolgov et~al., \emph{{On mass distribution of coalescing black holes}},
  \href{http://dx.doi.org/10.1088/1475-7516/2020/12/017}{\emph{JCAP} {\bf 12}
  (2020) 017}, [\href{http://arxiv.org/abs/2005.00892}{{\tt 2005.00892}}].

\bibitem{Hall:2020daa}
A.~Hall, A.~D. Gow and C.~T. Byrnes, \emph{{Bayesian analysis of LIGO-Virgo
  mergers: primordial vs. astrophysical black hole populations}},
  \href{http://dx.doi.org/10.1103/PhysRevD.102.123524}{\emph{Phys. Rev. D} {\bf
  102} (2020) 123524}, [\href{http://arxiv.org/abs/2008.13704}{{\tt
  2008.13704}}].

\bibitem{Hutsi:2020sol}
G.~H\"utsi, M.~Raidal, V.~Vaskonen and H.~Veerm\"ae, \emph{{Two populations of
  LIGO-Virgo black holes}},
  \href{http://dx.doi.org/10.1088/1475-7516/2021/03/068}{\emph{JCAP} {\bf 03}
  (2021) 068}, [\href{http://arxiv.org/abs/2012.02786}{{\tt 2012.02786}}].

\bibitem{1939PhRv...55..364T}
R.~C. {Tolman}, \emph{{Static solutions of Einstein's field equations for
  spheres of fluid}},
  \href{http://dx.doi.org/10.1103/PhysRev.55.364}{\emph{Physical Review} {\bf
  55} (Feb., 1939) 364--373}.

\bibitem{1939PhRv...55..374O}
J.~R. {Oppenheimer} and G.~M. {Volkoff}, \emph{{On massive neutron cores}},
  \href{http://dx.doi.org/10.1103/PhysRev.55.374}{\emph{Physical Review} {\bf
  55} (Feb., 1939) 374--381}.

\bibitem{Abbott:2020tfl}
{\scshape LIGO Scientific, Virgo} collaboration, R.~Abbott et~al.,
  \emph{{GW190521: a binary black hole merger with a total mass of $150
  M_{\odot}$}},
  \href{http://dx.doi.org/10.1103/PhysRevLett.125.101102}{\emph{Phys. Rev.
  Lett.} {\bf 125} (2020) 101102}, [\href{http://arxiv.org/abs/2009.01075}{{\tt
  2009.01075}}].

\bibitem{2018MNRAS.478.3756C}
B.~{Carr} and J.~{Silk}, \emph{{Primordial black holes as generators of cosmic
  structures}}, \href{http://dx.doi.org/10.1093/mnras/sty1204}{\emph{\mnras}
  {\bf 478} (Aug., 2018) 3756--3775},
  [\href{http://arxiv.org/abs/1801.00672}{{\tt 1801.00672}}].

\bibitem{Volonteri_2010}
M.~Volonteri, \emph{Formation of supermassive black holes},
  \href{http://dx.doi.org/10.1007/s00159-010-0029-x}{\emph{The Astronomy and
  Astrophysics Review} {\bf 18} (Apr, 2010) 279–315},
  [\href{http://arxiv.org/abs/1003.4404}{{\tt 1003.4404}}].

\bibitem{2016JCAP...02..064G}
J.~{Garriga}, A.~{Vilenkin} and J.~{Zhang}, \emph{{Black holes and the
  multiverse}},
  \href{http://dx.doi.org/10.1088/1475-7516/2016/02/064}{\emph{\jcap} {\bf
  2016} (Feb., 2016) 064}, [\href{http://arxiv.org/abs/1512.01819}{{\tt
  1512.01819}}].

\bibitem{2017JCAP...04..050D}
H.~{Deng}, J.~{Garriga} and A.~{Vilenkin}, \emph{{Primordial black hole and
  wormhole formation by domain walls}},
  \href{http://dx.doi.org/10.1088/1475-7516/2017/04/050}{\emph{\jcap} {\bf
  2017} (Apr., 2017) 050}, [\href{http://arxiv.org/abs/1612.03753}{{\tt
  1612.03753}}].

\bibitem{Hawking:1987bn}
S.~Hawking, \emph{{Black holes from cosmic strings}},
  \href{http://dx.doi.org/10.1016/0370-2693(89)90206-2}{\emph{Phys. Lett. B}
  {\bf 231} (1989) 237--239}.

\bibitem{PhysRevD.43.1106}
A.~Polnarev and R.~Zembowicz, \emph{Formation of primordial black holes by
  cosmic strings},
  \href{http://dx.doi.org/10.1103/PhysRevD.43.1106}{\emph{Phys. Rev. D} {\bf
  43} (Feb, 1991) 1106--1109}.

\bibitem{PhysRevD.26.2681}
S.~W. Hawking, I.~G. Moss and J.~M. Stewart, \emph{Bubble collisions in the
  very early universe},
  \href{http://dx.doi.org/10.1103/PhysRevD.26.2681}{\emph{Phys. Rev. D} {\bf
  26} (Nov, 1982) 2681--2693}.

\bibitem{2017JCAP...12..044D}
H.~{Deng} and A.~{Vilenkin}, \emph{{Primordial black hole formation by vacuum
  bubbles}},
  \href{http://dx.doi.org/10.1088/1475-7516/2017/12/044}{\emph{\jcap} {\bf
  2017} (Dec., 2017) 044}, [\href{http://arxiv.org/abs/1710.02865}{{\tt
  1710.02865}}].

\bibitem{Polnarev:1986bi}
A.~G. Polnarev and M.~Y. Khlopov, \emph{{Cosmology, primordial black holes, and
  supermassive particles}},
  \href{http://dx.doi.org/10.1070/PU1985v028n03ABEH003858}{\emph{Sov. Phys.
  Usp.} {\bf 28} (1985) 213--232}.

\bibitem{Khlopov:2008qy}
M.~Y. Khlopov, \emph{{Primordial black holes}},
  \href{http://dx.doi.org/10.1088/1674-4527/10/6/001}{\emph{Res. Astron.
  Astrophys.} {\bf 10} (2010) 495--528},
  [\href{http://arxiv.org/abs/0801.0116}{{\tt 0801.0116}}].

\bibitem{Belotsky:2014kca}
K.~M. Belotsky et~al., \emph{{Signatures of primordial black hole dark
  matter}}, \href{http://dx.doi.org/10.1142/S0217732314400057}{\emph{Mod. Phys.
  Lett. A} {\bf 29} (2014) 1440005},
  [\href{http://arxiv.org/abs/1410.0203}{{\tt 1410.0203}}].

\bibitem{Ketov:2019mfc}
S.~V. Ketov and M.~Y. Khlopov, \emph{{Cosmological probes of supersymmetric
  field theory models at superhigh energy scales}},
  \href{http://dx.doi.org/10.3390/sym11040511}{\emph{Symmetry} {\bf 11} (2019)
  511}.

\bibitem{1975ApJ...201....1C}
B.~J. {Carr}, \emph{{The primordial black hole mass spectrum.}},
  \href{http://dx.doi.org/10.1086/153853}{\emph{\apj} {\bf 201} (Oct., 1975)
  1--19}.

\bibitem{Niemeyer:1997mt}
J.~C. Niemeyer and K.~Jedamzik, \emph{{Near-critical gravitational collapse and
  the initial mass function of primordial black holes}},
  \href{http://dx.doi.org/10.1103/PhysRevLett.80.5481}{\emph{Phys. Rev. Lett.}
  {\bf 80} (1998) 5481--5484},
  [\href{http://arxiv.org/abs/astro-ph/9709072}{{\tt astro-ph/9709072}}].

\bibitem{Niemeyer:1999ak}
J.~C. Niemeyer and K.~Jedamzik, \emph{{Dynamics of primordial black hole
  formation}}, \href{http://dx.doi.org/10.1103/PhysRevD.59.124013}{\emph{Phys.
  Rev. D} {\bf 59} (1999) 124013},
  [\href{http://arxiv.org/abs/astro-ph/9901292}{{\tt astro-ph/9901292}}].

\bibitem{Musco:2004ak}
I.~Musco, J.~C. Miller and L.~Rezzolla, \emph{{Computations of primordial black
  hole formation}},
  \href{http://dx.doi.org/10.1088/0264-9381/22/7/013}{\emph{Class. Quant.
  Grav.} {\bf 22} (2005) 1405--1424},
  [\href{http://arxiv.org/abs/gr-qc/0412063}{{\tt gr-qc/0412063}}].

\bibitem{Escriva:2020tak}
A.~Escriv\`a, C.~Germani and R.~K. Sheth, \emph{{Analytical thresholds for
  black hole formation in general cosmological backgrounds}},
  \href{http://dx.doi.org/10.1088/1475-7516/2021/01/030}{\emph{JCAP} {\bf 01}
  (2021) 030}, [\href{http://arxiv.org/abs/2007.05564}{{\tt 2007.05564}}].

\bibitem{Musco:2020jjb}
I.~Musco, V.~De~Luca, G.~Franciolini and A.~Riotto, \emph{{Threshold for
  primordial black holes. II. A simple analytic prescription}},
  \href{http://dx.doi.org/10.1103/PhysRevD.103.063538}{\emph{Phys. Rev. D} {\bf
  103} (2021) 063538}, [\href{http://arxiv.org/abs/2011.03014}{{\tt
  2011.03014}}].

\bibitem{2014arXiv1403.1198G}
A.~M. {Green}, \emph{{Primordial black holes: sirens of the early Universe}},
  {\emph{arXiv e-prints} (Mar., 2014) arXiv:1403.1198},
  [\href{http://arxiv.org/abs/1403.1198}{{\tt 1403.1198}}].

\bibitem{2018JCAP...03..016F}
G.~{Franciolini}, A.~{Kehagias}, S.~{Matarrese} and A.~{Riotto},
  \emph{{Primordial black holes from inflation and non-Gaussianity}},
  \href{http://dx.doi.org/10.1088/1475-7516/2018/03/016}{\emph{\jcap} {\bf
  2018} (Mar., 2018) 016}, [\href{http://arxiv.org/abs/1801.09415}{{\tt
  1801.09415}}].

\bibitem{2019JCAP...07..048D}
V.~{De Luca}, G.~{Franciolini}, A.~{Kehagias}, M.~{Peloso}, A.~{Riotto} and
  C.~{{\"U}nal}, \emph{{The ineludible non-Gaussianity of the primordial black
  hole abundance}},
  \href{http://dx.doi.org/10.1088/1475-7516/2019/07/048}{\emph{\jcap} {\bf
  2019} (July, 2019) 048}, [\href{http://arxiv.org/abs/1904.00970}{{\tt
  1904.00970}}].

\bibitem{Atal:2018neu}
V.~Atal and C.~Germani, \emph{{The role of non-gaussianities in primordial
  black hole formation}},
  \href{http://dx.doi.org/10.1016/j.dark.2019.100275}{\emph{Phys. Dark Univ.}
  {\bf 24} (2019) 100275}, [\href{http://arxiv.org/abs/1811.07857}{{\tt
  1811.07857}}].

\bibitem{Cai:2018dig}
R.-g. Cai, S.~Pi and M.~Sasaki, \emph{{Gravitational waves induced by
  non-Gaussian scalar perturbations}},
  \href{http://dx.doi.org/10.1103/PhysRevLett.122.201101}{\emph{Phys. Rev.
  Lett.} {\bf 122} (2019) 201101}, [\href{http://arxiv.org/abs/1810.11000}{{\tt
  1810.11000}}].

\bibitem{Green:2004wb}
A.~M. Green, A.~R. Liddle, K.~A. Malik and M.~Sasaki, \emph{{A new calculation
  of the mass fraction of primordial black holes}},
  \href{http://dx.doi.org/10.1103/PhysRevD.70.041502}{\emph{Phys. Rev. D} {\bf
  70} (2004) 041502}, [\href{http://arxiv.org/abs/astro-ph/0403181}{{\tt
  astro-ph/0403181}}].

\bibitem{Germani:2018jgr}
C.~Germani and I.~Musco, \emph{{Abundance of primordial black holes depends on
  the shape of the inflationary power spectrum}},
  \href{http://dx.doi.org/10.1103/PhysRevLett.122.141302}{\emph{Phys. Rev.
  Lett.} {\bf 122} (2019) 141302}, [\href{http://arxiv.org/abs/1805.04087}{{\tt
  1805.04087}}].

\bibitem{Germani:2019zez}
C.~Germani and R.~K. Sheth, \emph{{Nonlinear statistics of primordial black
  holes from Gaussian curvature perturbations}},
  \href{http://dx.doi.org/10.1103/PhysRevD.101.063520}{\emph{Phys. Rev. D} {\bf
  101} (2020) 063520}, [\href{http://arxiv.org/abs/1912.07072}{{\tt
  1912.07072}}].

\bibitem{2010PhRvD..81j4019C}
B.~J. {Carr}, K.~{Kohri}, Y.~{Sendouda} and J.~{Yokoyama}, \emph{{New
  cosmological constraints on primordial black holes}},
  \href{http://dx.doi.org/10.1103/PhysRevD.81.104019}{\emph{\prd} {\bf 81}
  (May, 2010) 104019}, [\href{http://arxiv.org/abs/0912.5297}{{\tt
  0912.5297}}].

\bibitem{PhysRevD.58.083510}
J.~Yokoyama, \emph{Chaotic new inflation and formation of primordial black
  holes}, \href{http://dx.doi.org/10.1103/PhysRevD.58.083510}{\emph{Phys. Rev.
  D} {\bf 58} (Sep, 1998) 083510}.

\bibitem{2008JCAP...06..024S}
R.~{Saito}, J.~{Yokoyama} and R.~{Nagata}, \emph{{Single-field inflation,
  anomalous enhancement of superhorizon fluctuations and non-Gaussianity in
  primordial black hole formation}},
  \href{http://dx.doi.org/10.1088/1475-7516/2008/06/024}{\emph{\jcap} {\bf
  2008} (June, 2008) 024}, [\href{http://arxiv.org/abs/0804.3470}{{\tt
  0804.3470}}].

\bibitem{2017PDU....18...47G}
J.~{Garc{\'\i}a-Bellido} and E.~{Ruiz Morales}, \emph{{Primordial black holes
  from single field models of inflation}},
  \href{http://dx.doi.org/10.1016/j.dark.2017.09.007}{\emph{Physics of the Dark
  Universe} {\bf 18} (Dec., 2017) 47--54},
  [\href{http://arxiv.org/abs/1702.03901}{{\tt 1702.03901}}].

\bibitem{2017JPhCS.840a2032G}
J.~{Garc{\'\i}a-Bellido}, \emph{{Massive primordial black holes as dark matter
  and their detection with gravitational waves}},  in \emph{Journal of Physics
  Conference Series}, vol.~840 of \emph{Journal of Physics Conference Series},
  p.~012032, May, 2017.
\newblock \href{http://arxiv.org/abs/1702.08275}{{\tt 1702.08275}}.
\newblock \href{http://dx.doi.org/10.1088/1742-6596/840/1/012032}{DOI}.

\bibitem{2015PhRvD..92b3524C}
S.~{Clesse} and J.~{Garc{\'\i}a-Bellido}, \emph{{Massive primordial black holes
  from hybrid inflation as dark matter and the seeds of galaxies}},
  \href{http://dx.doi.org/10.1103/PhysRevD.92.023524}{\emph{\prd} {\bf 92}
  (July, 2015) 023524}, [\href{http://arxiv.org/abs/1501.07565}{{\tt
  1501.07565}}].

\bibitem{2017PhRvD..96b3514C}
B.~{Carr}, M.~{Raidal}, T.~{Tenkanen}, V.~{Vaskonen} and H.~{Veerm{\"a}e},
  \emph{{Primordial black hole constraints for extended mass functions}},
  \href{http://dx.doi.org/10.1103/PhysRevD.96.023514}{\emph{\prd} {\bf 96}
  (July, 2017) 023514}, [\href{http://arxiv.org/abs/1705.05567}{{\tt
  1705.05567}}].

\bibitem{2018JCAP...01..004B}
N.~{Bellomo}, J.~L. {Bernal}, A.~{Raccanelli} and L.~{Verde}, \emph{{Primordial
  black holes as dark matter: converting constraints from monochromatic to
  extended mass distributions}},
  \href{http://dx.doi.org/10.1088/1475-7516/2018/01/004}{\emph{\jcap} {\bf
  2018} (Jan., 2018) 004}, [\href{http://arxiv.org/abs/1709.07467}{{\tt
  1709.07467}}].

\bibitem{Lehmann:2018ejc}
B.~V. Lehmann, S.~Profumo and J.~Yant, \emph{{The maximal-density mass function
  for primordial black hole dark matter}},
  \href{http://dx.doi.org/10.1088/1475-7516/2018/04/007}{\emph{JCAP} {\bf 04}
  (2018) 007}, [\href{http://arxiv.org/abs/1801.00808}{{\tt 1801.00808}}].

\bibitem{Sureda:2020vgi}
J.~Sureda, J.~Maga\~na, I.~J. Araya and N.~D. Padilla, \emph{{Press-Schechter
  primordial black hole mass functions and their observational constraints}},
  \href{http://arxiv.org/abs/2008.09683}{{\tt 2008.09683}}.

\bibitem{Ashoorioon:2020hln}
A.~Ashoorioon, A.~Rostami and J.~T. Firouzjaee, \emph{{Charting the landscape
  in our neighborhood from the PBHs mass distribution and GWs}},
  \href{http://arxiv.org/abs/2012.02817}{{\tt 2012.02817}}.

\bibitem{1952MNRAS.112..195B}
H.~{Bondi}, \emph{{On spherically symmetrical accretion}},
  \href{http://dx.doi.org/10.1093/mnras/112.2.195}{\emph{\mnras} {\bf 112}
  (Jan., 1952) 195}.

\bibitem{Ali-Haimoud:2016mbv}
Y.~Ali-Ha{\"{\i}}moud and M.~Kamionkowski, \emph{{Cosmic microwave background
  limits on accreting primordial black holes}},
  \href{http://dx.doi.org/10.1103/PhysRevD.95.043534}{\emph{Phys. Rev.} {\bf
  D95} (2017) 043534}, [\href{http://arxiv.org/abs/1612.05644}{{\tt
  1612.05644}}].

\bibitem{Agol:2001hb}
E.~Agol and M.~Kamionkowski, \emph{{X-rays from isolated black holes in the
  Milky Way}},
  \href{http://dx.doi.org/10.1046/j.1365-8711.2002.05523.x}{\emph{Mon. Not.
  Roy. Astron. Soc.} {\bf 334} (2002) 553},
  [\href{http://arxiv.org/abs/astro-ph/0109539}{{\tt astro-ph/0109539}}].

\bibitem{Poulin:2017bwe}
V.~Poulin, P.~D. Serpico, F.~Calore, S.~Clesse and K.~Kohri, \emph{{CMB bounds
  on disk-accreting massive primordial black holes}},
  \href{http://dx.doi.org/10.1103/PhysRevD.96.083524}{\emph{Phys. Rev.} {\bf
  D96} (2017) 083524}, [\href{http://arxiv.org/abs/1707.04206}{{\tt
  1707.04206}}].

\bibitem{Xie:2012rs}
F.-G. Xie and F.~Yuan, \emph{{The radiative efficiency of hot accretion
  flows}}, \href{http://dx.doi.org/10.1111/j.1365-2966.2012.22030.x}{\emph{Mon.
  Not. Roy. Astron. Soc.} {\bf 427} (2012) 1580},
  [\href{http://arxiv.org/abs/1207.3113}{{\tt 1207.3113}}].

\bibitem{Shakura:1972te}
N.~I. Shakura and R.~A. Sunyaev, \emph{{Black holes in binary systems.
  Observational appearance}}, {\emph{Astron. Astrophys.} {\bf 24} (1973)
  337--355}.

\bibitem{Ichimaru:1977uf}
S.~Ichimaru, \emph{{Bimodal behavior of accretion disks - Theory and
  application to Cygnus X-1 transitions}},
  \href{http://dx.doi.org/10.1086/155314}{\emph{Astrophys. J.} {\bf 214} (1977)
  840--855}.

\bibitem{Rees:1982pe}
M.~J. Rees, E.~S. Phinney, M.~C. Begelman and R.~D. Blandford, \emph{{Ion
  supported tori and the origin of radio jets}},
  \href{http://dx.doi.org/10.1038/295017a0}{\emph{Nature} {\bf 295} (1982)
  17--21}.

\bibitem{1994ApJ...428L..13N}
R.~{Narayan} and I.~{Yi}, \emph{{Advection-dominated accretion: a self-similar
  solution}}, \href{http://dx.doi.org/10.1086/187381}{\emph{\apjl} {\bf 428}
  (June, 1994) L13}, [\href{http://arxiv.org/abs/astro-ph/9403052}{{\tt
  astro-ph/9403052}}].

\bibitem{Narayan:1994is}
R.~Narayan and I.~Yi, \emph{{Advection dominated accretion: underfed black
  holes and neutron stars}},
  \href{http://dx.doi.org/10.1086/176343}{\emph{Astrophys. J.} {\bf 452} (1995)
  710}, [\href{http://arxiv.org/abs/astro-ph/9411059}{{\tt astro-ph/9411059}}].

\bibitem{Yuan:2014gma}
F.~Yuan and R.~Narayan, \emph{{Hot accretion flows around black holes}},
  \href{http://dx.doi.org/10.1146/annurev-astro-082812-141003}{\emph{Ann. Rev.
  Astron. Astrophys.} {\bf 52} (2014) 529--588},
  [\href{http://arxiv.org/abs/1401.0586}{{\tt 1401.0586}}].

\bibitem{Slatyer:2015kla}
T.~R. Slatyer, \emph{{Indirect dark matter signatures in the cosmic dark ages
  II. Ionization, heating and photon production from arbitrary energy
  injections}}, \href{http://dx.doi.org/10.1103/PhysRevD.93.023521}{\emph{Phys.
  Rev.} {\bf D93} (2016) 023521}, [\href{http://arxiv.org/abs/1506.03812}{{\tt
  1506.03812}}].

\bibitem{1974Natur.248...30H}
S.~W. {Hawking}, \emph{{Black hole explosions?}},
  \href{http://dx.doi.org/10.1038/248030a0}{\emph{\nat} {\bf 248} (Mar., 1974)
  30--31}.

\bibitem{2003PhTea..41..299L}
M.~C. {Lopresto}, \emph{{Some simple black hole thermodynamics}},
  \href{http://dx.doi.org/10.1119/1.1571268}{\emph{The Physics Teacher} {\bf
  41} (May, 2003) 299--301}.

\bibitem{1976PhRvD..14.3260P}
D.~N. {Page}, \emph{{Particle emission rates from a black hole. II. Massless
  particles from a rotating hole}},
  \href{http://dx.doi.org/10.1103/PhysRevD.14.3260}{\emph{\prd} {\bf 14} (Dec.,
  1976) 3260--3273}.

\bibitem{Arbey:2019mbc}
A.~Arbey and J.~Auffinger, \emph{{BlackHawk: a public code for calculating the
  Hawking evaporation spectra of any black hole distribution}},
  \href{http://dx.doi.org/10.1140/epjc/s10052-019-7161-1}{\emph{Eur. Phys. J.
  C} {\bf 79} (2019) 693}, [\href{http://arxiv.org/abs/1905.04268}{{\tt
  1905.04268}}].

\bibitem{2018PhRvL.121h1304A}
Y.~{Ali-Ha{\"\i}moud}, \emph{{Correlation function of high-threshold regions
  and application to the initial small-scale clustering of primordial black
  holes}}, \href{http://dx.doi.org/10.1103/PhysRevLett.121.081304}{\emph{\prl}
  {\bf 121} (Aug., 2018) 081304}, [\href{http://arxiv.org/abs/1805.05912}{{\tt
  1805.05912}}].

\bibitem{2018PhRvD..98l3533D}
V.~{Desjacques} and A.~{Riotto}, \emph{{Spatial clustering of primordial black
  holes}}, \href{http://dx.doi.org/10.1103/PhysRevD.98.123533}{\emph{\prd} {\bf
  98} (Dec., 2018) 123533}, [\href{http://arxiv.org/abs/1806.10414}{{\tt
  1806.10414}}].

\bibitem{2019PTEP.2019j3E02S}
T.~{Suyama} and S.~{Yokoyama}, \emph{{Clustering of primordial black holes with
  non-Gaussian initial fluctuations}},
  \href{http://dx.doi.org/10.1093/ptep/ptz105}{\emph{Progress of Theoretical
  and Experimental Physics} {\bf 2019} (Oct., 2019) 103E02},
  [\href{http://arxiv.org/abs/1906.04958}{{\tt 1906.04958}}].

\bibitem{2018JCAP...10..043B}
G.~{Ballesteros}, P.~D. {Serpico} and M.~{Taoso}, \emph{{On the merger rate of
  primordial black holes: effects of nearest neighbours distribution and
  clustering}},
  \href{http://dx.doi.org/10.1088/1475-7516/2018/10/043}{\emph{\jcap} {\bf
  2018} (Oct., 2018) 043}, [\href{http://arxiv.org/abs/1807.02084}{{\tt
  1807.02084}}].

\bibitem{2019JCAP...11..001M}
A.~{Moradinezhad Dizgah}, G.~{Franciolini} and A.~{Riotto}, \emph{{Primordial
  black holes from broad spectra: abundance and clustering}},
  \href{http://dx.doi.org/10.1088/1475-7516/2019/11/001}{\emph{\jcap} {\bf
  2019} (Nov., 2019) 001}, [\href{http://arxiv.org/abs/1906.08978}{{\tt
  1906.08978}}].

\bibitem{DeLuca:2020jug}
V.~De~Luca, V.~Desjacques, G.~Franciolini and A.~Riotto, \emph{{The clustering
  evolution of primordial black holes}},
  \href{http://dx.doi.org/10.1088/1475-7516/2020/11/028}{\emph{JCAP} {\bf 11}
  (2020) 028}, [\href{http://arxiv.org/abs/2009.04731}{{\tt 2009.04731}}].

\bibitem{Belotsky:2018wph}
K.~M. Belotsky et~al., \emph{{Clusters of primordial black holes}},
  \href{http://dx.doi.org/10.1140/epjc/s10052-019-6741-4}{\emph{Eur. Phys. J.
  C} {\bf 79} (2019) 246}, [\href{http://arxiv.org/abs/1807.06590}{{\tt
  1807.06590}}].

\bibitem{Chiba:2017rvs}
T.~Chiba and S.~Yokoyama, \emph{{Spin distribution of primordial black holes}},
  \href{http://dx.doi.org/10.1093/ptep/ptx087}{\emph{PTEP} {\bf 2017} (2017)
  083E01}, [\href{http://arxiv.org/abs/1704.06573}{{\tt 1704.06573}}].

\bibitem{Mirbabayi:2019uph}
M.~Mirbabayi, A.~Gruzinov and J.~Nore\~na, \emph{{Spin of primordial black
  holes}}, \href{http://dx.doi.org/10.1088/1475-7516/2020/03/017}{\emph{JCAP}
  {\bf 03} (2020) 017}, [\href{http://arxiv.org/abs/1901.05963}{{\tt
  1901.05963}}].

\bibitem{2019JCAP...05..018D}
V.~{De Luca}, V.~{Desjacques}, G.~{Franciolini}, A.~{Malhotra} and A.~{Riotto},
  \emph{{The initial spin probability distribution of primordial black holes}},
  \href{http://dx.doi.org/10.1088/1475-7516/2019/05/018}{\emph{\jcap} {\bf
  2019} (May, 2019) 018}, [\href{http://arxiv.org/abs/1903.01179}{{\tt
  1903.01179}}].

\bibitem{DeLuca:2020bjf}
V.~De~Luca, G.~Franciolini, P.~Pani and A.~Riotto, \emph{{The evolution of
  primordial black holes and their final observable spins}},
  \href{http://dx.doi.org/10.1088/1475-7516/2020/04/052}{\emph{JCAP} {\bf 04}
  (2020) 052}, [\href{http://arxiv.org/abs/2003.02778}{{\tt 2003.02778}}].

\bibitem{2003ApJ...594L..71A}
N.~{Afshordi}, P.~{McDonald} and D.~N. {Spergel}, \emph{{Primordial black holes
  as dark matter: the power spectrum and evaporation of early structures}},
  \href{http://dx.doi.org/10.1086/378763}{\emph{\apjl} {\bf 594} (Sept., 2003)
  L71--L74}, [\href{http://arxiv.org/abs/astro-ph/0302035}{{\tt
  astro-ph/0302035}}].

\bibitem{Gong:2017sie}
J.-O. Gong and N.~Kitajima, \emph{{Small-scale structure and 21cm fluctuations
  by primordial black holes}},
  \href{http://dx.doi.org/10.1088/1475-7516/2017/08/017}{\emph{JCAP} {\bf 1708}
  (2017) 017}, [\href{http://arxiv.org/abs/1704.04132}{{\tt 1704.04132}}].

\bibitem{2006PhRvD..73h3504C}
J.~R. {Chisholm}, \emph{{Clustering of primordial black holes: basic results}},
  \href{http://dx.doi.org/10.1103/PhysRevD.73.083504}{\emph{\prd} {\bf 73}
  (Apr., 2006) 083504}, [\href{http://arxiv.org/abs/astro-ph/0509141}{{\tt
  astro-ph/0509141}}].

\bibitem{Inman:2019wvr}
D.~Inman and Y.~Ali-Ha\"\i{}moud, \emph{{Early structure formation in
  primordial black hole cosmologies}},
  \href{http://dx.doi.org/10.1103/PhysRevD.100.083528}{\emph{Phys. Rev. D} {\bf
  100} (2019) 083528}, [\href{http://arxiv.org/abs/1907.08129}{{\tt
  1907.08129}}].

\bibitem{Peacock:1999ye}
J.~A. Peacock, \emph{{Cosmological physics}}.
\newblock Cambridge University Press, 1999.
\newblock 10.1017/CBO9780511804533.

\bibitem{Gong:2018sos}
J.-O. Gong and N.~Kitajima, \emph{{Distribution of primordial black holes and
  21cm signature}},
  \href{http://dx.doi.org/10.1088/1475-7516/2018/11/041}{\emph{JCAP} {\bf 1811}
  (2018) 041}, [\href{http://arxiv.org/abs/1803.02745}{{\tt 1803.02745}}].

\bibitem{Mena:2019nhm}
O.~Mena, S.~Palomares-Ruiz, P.~Villanueva-Domingo and S.~J. Witte,
  \emph{{Constraining the primordial black hole abundance with 21-cm
  cosmology}}, \href{http://dx.doi.org/10.1103/PhysRevD.100.043540}{\emph{Phys.
  Rev.} {\bf D100} (2019) 043540}, [\href{http://arxiv.org/abs/1906.07735}{{\tt
  1906.07735}}].

\bibitem{Clark:2016nst}
S.~Clark, B.~Dutta, Y.~Gao, L.~E. Strigari and S.~Watson, \emph{{Planck
  constraint on relic primordial black holes}},
  \href{http://dx.doi.org/10.1103/PhysRevD.95.083006}{\emph{Phys. Rev. D} {\bf
  95} (2017) 083006}, [\href{http://arxiv.org/abs/1612.07738}{{\tt
  1612.07738}}].

\bibitem{Alcock:2000kg}
{\scshape Macho} collaboration, C.~Alcock et~al., \emph{{MACHO project limits
  on black hole dark matter in the 1-30 M$_\odot$ range}},
  \href{http://dx.doi.org/10.1086/319636}{\emph{Astrophys. J. Lett.} {\bf 550}
  (2001) L169}, [\href{http://arxiv.org/abs/astro-ph/0011506}{{\tt
  astro-ph/0011506}}].

\bibitem{Tisserand:2006zx}
{\scshape EROS-2} collaboration, P.~Tisserand et~al., \emph{{Limits on the
  MACHO content of the galactic halo from the EROS-2 survey of the Magellanic
  Clouds}}, \href{http://dx.doi.org/10.1051/0004-6361:20066017}{\emph{Astron.
  Astrophys.} {\bf 469} (2007) 387--404},
  [\href{http://arxiv.org/abs/astro-ph/0607207}{{\tt astro-ph/0607207}}].

\bibitem{Griest:2013aaa}
K.~Griest, A.~M. Cieplak and M.~J. Lehner, \emph{{Experimental limits on
  primordial black hole dark matter from the first 2 yr of Kepler data}},
  \href{http://dx.doi.org/10.1088/0004-637X/786/2/158}{\emph{Astrophys. J.}
  {\bf 786} (2014) 158}, [\href{http://arxiv.org/abs/1307.5798}{{\tt
  1307.5798}}].

\bibitem{Oguri:2017ock}
M.~Oguri, J.~M. Diego, N.~Kaiser, P.~L. Kelly and T.~Broadhurst,
  \emph{{Understanding caustic crossings in giant arcs: characteristic scales,
  event rates, and constraints on compact dark matter}},
  \href{http://dx.doi.org/10.1103/PhysRevD.97.023518}{\emph{Phys. Rev. D} {\bf
  97} (2018) 023518}, [\href{http://arxiv.org/abs/1710.00148}{{\tt
  1710.00148}}].

\bibitem{2019PhRvD..99h3503N}
H.~{Niikura}, M.~{Takada}, S.~{Yokoyama}, T.~{Sumi} and S.~{Masaki},
  \emph{{Constraints on Earth-mass primordial black holes from OGLE 5-year
  microlensing events}},
  \href{http://dx.doi.org/10.1103/PhysRevD.99.083503}{\emph{\prd} {\bf 99}
  (Apr., 2019) 083503}, [\href{http://arxiv.org/abs/1901.07120}{{\tt
  1901.07120}}].

\bibitem{Croon:2020ouk}
D.~Croon, D.~McKeen, N.~Raj and Z.~Wang, \emph{{Subaru-HSC through a different
  lens: microlensing by extended dark matter structures}},
  \href{http://dx.doi.org/10.1103/PhysRevD.102.083021}{\emph{Phys. Rev. D} {\bf
  102} (2020) 083021}, [\href{http://arxiv.org/abs/2007.12697}{{\tt
  2007.12697}}].

\bibitem{2020PhRvR...2b3204S}
P.~D. {Serpico}, V.~{Poulin}, D.~{Inman} and K.~{Kohri}, \emph{{Cosmic
  microwave background bounds on primordial black holes including dark matter
  halo accretion}},
  \href{http://dx.doi.org/10.1103/PhysRevResearch.2.023204}{\emph{Physical
  Review Research} {\bf 2} (May, 2020) 023204},
  [\href{http://arxiv.org/abs/2002.10771}{{\tt 2002.10771}}].

\bibitem{2014ApJ...790..159M}
M.~A. {Monroy-Rodr{\'\i}guez} and C.~{Allen}, \emph{{The end of the MACHO era
  revisited: new limits on MACHO masses from halo wide binaries}},
  \href{http://dx.doi.org/10.1088/0004-637X/790/2/159}{\emph{\apj} {\bf 790}
  (Aug., 2014) 159}, [\href{http://arxiv.org/abs/1406.5169}{{\tt 1406.5169}}].

\bibitem{2016ApJ...824L..31B}
T.~D. {Brandt}, \emph{{Constraints on MACHO dark matter from compact stellar
  systems in ultra-faint dwarf galaxies}},
  \href{http://dx.doi.org/10.3847/2041-8205/824/2/L31}{\emph{\apjl} {\bf 824}
  (June, 2016) L31}, [\href{http://arxiv.org/abs/1605.03665}{{\tt
  1605.03665}}].

\bibitem{2019PhRvL.123g1102M}
R.~{Murgia}, G.~{Scelfo}, M.~{Viel} and A.~{Raccanelli},
  \emph{{Lyman-{\ensuremath{\alpha}} forest constraints on primordial black
  holes as dark matter}},
  \href{http://dx.doi.org/10.1103/PhysRevLett.123.071102}{\emph{\prl} {\bf 123}
  (Aug., 2019) 071102}, [\href{http://arxiv.org/abs/1903.10509}{{\tt
  1903.10509}}].

\bibitem{2018PhRvD..98b3536K}
B.~J. {Kavanagh}, D.~{Gaggero} and G.~{Bertone}, \emph{{Merger rate of a
  subdominant population of primordial black holes}},
  \href{http://dx.doi.org/10.1103/PhysRevD.98.023536}{\emph{\prd} {\bf 98}
  (July, 2018) 023536}, [\href{http://arxiv.org/abs/1805.09034}{{\tt
  1805.09034}}].

\bibitem{Authors:2019qbw}
{\scshape LIGO Scientific, Virgo} collaboration, B.~P. Abbott et~al.,
  \emph{{Search for subsolar mass ultracompact binaries in advanced
  LIGO\textquoteright{}s second observing run}},
  \href{http://dx.doi.org/10.1103/PhysRevLett.123.161102}{\emph{Phys. Rev.
  Lett.} {\bf 123} (2019) 161102}, [\href{http://arxiv.org/abs/1904.08976}{{\tt
  1904.08976}}].

\bibitem{2020JCAP...08..039C}
Z.-C. {Chen} and Q.-G. {Huang}, \emph{{Distinguishing primordial black holes
  from astrophysical black holes by Einstein Telescope and Cosmic Explorer}},
  \href{http://dx.doi.org/10.1088/1475-7516/2020/08/039}{\emph{\jcap} {\bf
  2020} (Aug., 2020) 039}, [\href{http://arxiv.org/abs/1904.02396}{{\tt
  1904.02396}}].

\bibitem{Boehm:2020jwd}
C.~Boehm, A.~Kobakhidze, C.~A.~J. O'hare, Z.~S.~C. Picker and M.~Sakellariadou,
  \emph{{Eliminating the LIGO bounds on primordial black hole dark matter}},
  \href{http://dx.doi.org/10.1088/1475-7516/2021/03/078}{\emph{JCAP} {\bf 03}
  (2021) 078}, [\href{http://arxiv.org/abs/2008.10743}{{\tt 2008.10743}}].

\bibitem{Villanueva-Domingo:2021cgh}
P.~Villanueva-Domingo and K.~Ichiki, \emph{{21 cm forest constraints on
  primordial black holes}},  \href{http://arxiv.org/abs/2104.10695}{{\tt
  2104.10695}}.

\bibitem{bradley_j_kavanagh_2019_3538999}
B.~J. Kavanagh, \emph{bradkav/pbhbounds: Release version},  Nov., 2019.
\newblock 10.5281/zenodo.3538999.

\bibitem{2016PhRvD..94h3504C}
B.~{Carr}, F.~{K{\"u}hnel} and M.~{Sandstad}, \emph{{Primordial black holes as
  dark matter}},
  \href{http://dx.doi.org/10.1103/PhysRevD.94.083504}{\emph{\prd} {\bf 94}
  (Oct., 2016) 083504}, [\href{http://arxiv.org/abs/1607.06077}{{\tt
  1607.06077}}].

\bibitem{DeRocco:2019fjq}
W.~DeRocco and P.~W. Graham, \emph{{Constraining primordial black hole
  abundance with the galactic 511 keV line}},
  \href{http://dx.doi.org/10.1103/PhysRevLett.123.251102}{\emph{Phys. Rev.
  Lett.} {\bf 123} (2019) 251102}, [\href{http://arxiv.org/abs/1906.07740}{{\tt
  1906.07740}}].

\bibitem{Laha:2019ssq}
R.~Laha, \emph{{Primordial black holes as a dark matter candidate are severely
  constrained by the galactic center 511 keV $\gamma$ -ray line}},
  \href{http://dx.doi.org/10.1103/PhysRevLett.123.251101}{\emph{Phys. Rev.
  Lett.} {\bf 123} (2019) 251101}, [\href{http://arxiv.org/abs/1906.09994}{{\tt
  1906.09994}}].

\bibitem{Laha:2020ivk}
R.~Laha, J.~B. Mu\~noz and T.~R. Slatyer, \emph{{INTEGRAL constraints on
  primordial black holes and particle dark matter}},
  \href{http://dx.doi.org/10.1103/PhysRevD.101.123514}{\emph{Phys. Rev. D} {\bf
  101} (2020) 123514}, [\href{http://arxiv.org/abs/2004.00627}{{\tt
  2004.00627}}].

\bibitem{Coogan:2020tuf}
A.~Coogan, L.~Morrison and S.~Profumo, \emph{{Direct detection of Hawking
  radiation from asteroid-mass primordial black holes}},
  \href{http://arxiv.org/abs/2010.04797}{{\tt 2010.04797}}.

\bibitem{Ballesteros:2019exr}
G.~Ballesteros, J.~Coronado-Bl\'azquez and D.~Gaggero, \emph{{X-ray and
  gamma-ray limits on the primordial black hole abundance from Hawking
  radiation}},
  \href{http://dx.doi.org/10.1016/j.physletb.2020.135624}{\emph{Phys. Lett. B}
  {\bf 808} (2020) 135624}, [\href{http://arxiv.org/abs/1906.10113}{{\tt
  1906.10113}}].

\bibitem{Iguaz:2021irx}
J.~Iguaz, P.~D. Serpico and T.~Siegert, \emph{{Isotropic X-ray bound on
  primordial black hole dark matter}},
  \href{http://arxiv.org/abs/2104.03145}{{\tt 2104.03145}}.

\bibitem{Dasgupta:2019cae}
B.~Dasgupta, R.~Laha and A.~Ray, \emph{{Neutrino and positron constraints on
  spinning primordial black hole dark matter}},
  \href{http://dx.doi.org/10.1103/PhysRevLett.125.101101}{\emph{Phys. Rev.
  Lett.} {\bf 125} (2020) 101101}, [\href{http://arxiv.org/abs/1912.01014}{{\tt
  1912.01014}}].

\bibitem{1986ApJ...304....1P}
B.~{Paczynski}, \emph{{Gravitational microlensing by the Galactic halo}},
  \href{http://dx.doi.org/10.1086/164140}{\emph{\apj} {\bf 304} (May, 1986) 1}.

\bibitem{2019NatAs...3..524N}
H.~{Niikura}, M.~{Takada}, N.~{Yasuda}, R.~H. {Lupton}, T.~{Sumi}, S.~{More}
  et~al., \emph{{Microlensing constraints on primordial black holes with
  Subaru/HSC Andromeda observations}},
  \href{http://dx.doi.org/10.1038/s41550-019-0723-1}{\emph{Nature Astronomy}
  {\bf 3} (Apr., 2019) 524--534}, [\href{http://arxiv.org/abs/1701.02151}{{\tt
  1701.02151}}].

\bibitem{2017Natur.548..183M}
P.~{Mr{\'o}z} et~al., \emph{{No large population of unbound or wide-orbit
  Jupiter-mass planets}},
  \href{http://dx.doi.org/10.1038/nature23276}{\emph{\nat} {\bf 548} (Aug.,
  2017) 183--186}, [\href{http://arxiv.org/abs/1707.07634}{{\tt 1707.07634}}].

\bibitem{2015A&A...575A.107H}
M.~R.~S. {Hawkins}, \emph{{A new look at microlensing limits on dark matter in
  the galactic halo}},
  \href{http://dx.doi.org/10.1051/0004-6361/201425400}{\emph{\aap} {\bf 575}
  (Mar., 2015) A107}, [\href{http://arxiv.org/abs/1503.01935}{{\tt
  1503.01935}}].

\bibitem{2018PhRvL.121n1101Z}
M.~{Zumalac{\'a}rregui} and U.~{Seljak}, \emph{{Limits on stellar-mass compact
  objects as dark matter from gravitational lensing of type Ia supernovae}},
  \href{http://dx.doi.org/10.1103/PhysRevLett.121.141101}{\emph{\prl} {\bf 121}
  (Oct., 2018) 141101}, [\href{http://arxiv.org/abs/1712.02240}{{\tt
  1712.02240}}].

\bibitem{2012PhRvD..86d3001B}
A.~{Barnacka}, J.~F. {Glicenstein} and R.~{Moderski}, \emph{{New constraints on
  primordial black holes abundance from femtolensing of gamma-ray bursts}},
  \href{http://dx.doi.org/10.1103/PhysRevD.86.043001}{\emph{\prd} {\bf 86}
  (Aug., 2012) 043001}, [\href{http://arxiv.org/abs/1204.2056}{{\tt
  1204.2056}}].

\bibitem{2017arXiv171206574G}
J.~{Garcia-Bellido}, S.~{Clesse} and P.~{Fleury}, \emph{{LIGO Lo(g)Normal
  MACHO: primordial black holes survive SN lensing constraints}}, {\emph{arXiv
  e-prints} (Dec., 2017) arXiv:1712.06574},
  [\href{http://arxiv.org/abs/1712.06574}{{\tt 1712.06574}}].

\bibitem{2018JCAP...12..005K}
A.~{Katz}, J.~{Kopp}, S.~{Sibiryakov} and W.~{Xue}, \emph{{Femtolensing by dark
  matter revisited}},
  \href{http://dx.doi.org/10.1088/1475-7516/2018/12/005}{\emph{\jcap} {\bf
  2018} (Dec., 2018) 005}, [\href{http://arxiv.org/abs/1807.11495}{{\tt
  1807.11495}}].

\bibitem{1981InJPh..55..304T}
S.~N.~G. {Thakurta}, \emph{{Kerr metric in an expanding universe}},
  {\emph{Indian Journal of Physics} {\bf 55B} (Jan., 1981) 304--310}.

\bibitem{2008ApJ...680..829R}
M.~{Ricotti}, J.~P. {Ostriker} and K.~J. {Mack}, \emph{{Effect of primordial
  black holes on the cosmic microwave background and cosmological parameter
  estimates}}, \href{http://dx.doi.org/10.1086/587831}{\emph{\apj} {\bf 680}
  (June, 2008) 829--845}, [\href{http://arxiv.org/abs/0709.0524}{{\tt
  0709.0524}}].

\bibitem{2016arXiv161207264H}
B.~{Horowitz}, \emph{{Revisiting primordial black holes constraints from
  ionization history}}, {\emph{arXiv e-prints} (Dec., 2016) arXiv:1612.07264},
  [\href{http://arxiv.org/abs/1612.07264}{{\tt 1612.07264}}].

\bibitem{Poulin:2016anj}
V.~Poulin, J.~Lesgourgues and P.~D. Serpico, \emph{{Cosmological constraints on
  exotic injection of electromagnetic energy}},
  \href{http://dx.doi.org/10.1088/1475-7516/2017/03/043}{\emph{JCAP} {\bf 03}
  (2017) 043}, [\href{http://arxiv.org/abs/1610.10051}{{\tt 1610.10051}}].

\bibitem{Acharya:2019xla}
S.~K. Acharya and R.~Khatri, \emph{{CMB spectral distortions constraints on
  primordial black holes, cosmic strings and long lived unstable particles
  revisited}},
  \href{http://dx.doi.org/10.1088/1475-7516/2020/02/010}{\emph{JCAP} {\bf 02}
  (2020) 010}, [\href{http://arxiv.org/abs/1912.10995}{{\tt 1912.10995}}].

\bibitem{Nakama:2017xvq}
T.~Nakama, B.~Carr and J.~Silk, \emph{{Limits on primordial black holes from
  $\mu$ distortions in cosmic microwave background}},
  \href{http://dx.doi.org/10.1103/PhysRevD.97.043525}{\emph{Phys. Rev. D} {\bf
  97} (2018) 043525}, [\href{http://arxiv.org/abs/1710.06945}{{\tt
  1710.06945}}].

\bibitem{Bowman:2018yin}
J.~D. Bowman, A.~E.~E. Rogers, R.~A. Monsalve, T.~J. Mozdzen and N.~Mahesh,
  \emph{{An absorption profile centred at 78 megahertz in the sky-averaged
  spectrum}}, \href{http://dx.doi.org/10.1038/nature25792}{\emph{Nature} {\bf
  555} (2018) 67--70}.

\bibitem{Hektor:2018qqw}
A.~Hektor, G.~H\"utsi, L.~Marzola, M.~Raidal, V.~Vaskonen and H.~Veerm\"ae,
  \emph{{Constraining primordial black holes with the EDGES 21-cm absorption
  signal}}, \href{http://dx.doi.org/10.1103/PhysRevD.98.023503}{\emph{Phys.
  Rev. D} {\bf 98} (2018) 023503}, [\href{http://arxiv.org/abs/1803.09697}{{\tt
  1803.09697}}].

\bibitem{Clark:2018ghm}
S.~Clark, B.~Dutta, Y.~Gao, Y.-Z. Ma and L.~E. Strigari, \emph{{21 cm limits on
  decaying dark matter and primordial black holes}},
  \href{http://dx.doi.org/10.1103/PhysRevD.98.043006}{\emph{Phys. Rev. D} {\bf
  98} (2018) 043006}, [\href{http://arxiv.org/abs/1803.09390}{{\tt
  1803.09390}}].

\bibitem{Halder:2021jiv}
A.~Halder and M.~Pandey, \emph{{Investigating the effect of PBH, dark matter --
  baryon and dark matter -- dark energy interaction on EDGES in 21cm signal}},
  \href{http://arxiv.org/abs/2101.05228}{{\tt 2101.05228}}.

\bibitem{Halder:2021rbq}
A.~Halder and S.~Banerjee, \emph{{Bounds on abundance of primordial black hole
  and dark matter from EDGES 21-cm signal}},
  \href{http://dx.doi.org/10.1103/PhysRevD.103.063044}{\emph{Phys. Rev. D} {\bf
  103} (2021) 063044}, [\href{http://arxiv.org/abs/2102.00959}{{\tt
  2102.00959}}].

\bibitem{2018Natur.564E..32H}
R.~{Hills}, G.~{Kulkarni}, P.~D. {Meerburg} and E.~{Puchwein}, \emph{{Concerns
  about modelling of the EDGES data}},
  \href{http://dx.doi.org/10.1038/s41586-018-0796-5}{\emph{\nat} {\bf 564}
  (Dec., 2018) E32--E34}, [\href{http://arxiv.org/abs/1805.01421}{{\tt
  1805.01421}}].

\bibitem{2019ApJ...874..153B}
R.~F. {Bradley}, K.~{Tauscher}, D.~{Rapetti} and J.~O. {Burns}, \emph{{A ground
  plane artifact that induces an absorption profile in averaged spectra from
  global 21 cm measurements, with possible application to EDGES}},
  \href{http://dx.doi.org/10.3847/1538-4357/ab0d8b}{\emph{\apj} {\bf 874}
  (Apr., 2019) 153}, [\href{http://arxiv.org/abs/1810.09015}{{\tt
  1810.09015}}].

\bibitem{2020MNRAS.492...22S}
P.~H. {Sims} and J.~C. {Pober}, \emph{{Testing for calibration systematics in
  the EDGES low-band data using Bayesian model selection}},
  \href{http://dx.doi.org/10.1093/mnras/stz3388}{\emph{\mnras} {\bf 492} (Feb.,
  2020) 22--38}, [\href{http://arxiv.org/abs/1910.03165}{{\tt 1910.03165}}].

\end{thebibliography}\endgroup

\end{document}